\documentclass[preprint,aps,tightenlines,floatfix]{revtex4}

\usepackage{graphicx}
\usepackage{bm}
\usepackage{amsmath}
\usepackage{amssymb}
\usepackage{enumerate}
\usepackage{mathrsfs}
\usepackage{yfonts}
\usepackage{textcomp}
\usepackage{units}

\DeclareMathAlphabet{\mathpzc}{OT1}{pzc}{m}{it}

\begin{document}

\title{A Brief Review Noether's Theorems  and their Application to General Relativity} 

\author{R. J. McLeod}
\email{robert.mcleod@flinders.edu.au}

\affiliation{
School of Chemical and Physical Sciences, Flinders University\\ Bedford Park, South Australia, 5042, Australia}

\date{\today}

\begin{abstract}

In this article, we will review Noether's Theorems and their application in General Relativity.
We will present Noether's Theorems in their original form and restate them as they are usually applied to physics.  
Some basic equations of Special Relativity will be reviewed and contrasted with the equations in General Relativity.  
We will be most interested in the role of Noether's Theorems in  conservation laws.  
Several applications in flat spaces are examined.  A formulation of Noether's Theorems  in curved space is presented.  Of particular interest is  the conservation of energy in General Relativity.  
Here, there is no general form for a conserved tensor, as there is in a flat space case.  
Pseudotensor formulations can be found and two examples are given.  
In special cases, one can formulate an equation for the conservation of energy.  
The standard cosmology is an example.  
 \\
 

\end{abstract}


\maketitle

 \tableofcontents

\newpage

%%%%%%%%%%%%%%%%%%%%%%%%%%%%%%%%%%%%%%%%%%%
%%%%%%%%%%%%%%%%%%%%%%%%%%%%%%%%%%%%%%%%%%%

\section{Introduction}
\label{sec:level1}

The main motivation for this work was to gain an understanding of the lack of a general formulation of the conservation of energy in General Relativity (GR).  
Conservation of energy is a basic concept in physics.  
Its formulation in mechanics and Special Relativity (SR), relies on the space-time being flat.  In GR, the space is no longer flat and the construction of energy conservation is not universal.  
The role that Noether's Theorems \cite{Noe1918} plays in our understanding is vital 
since they supply the mathematics to understand this difficulty.  
As such, this review will centre on developing an understanding of Noether's Theorems and the role they play in physics.  Much of the presentation of Noether's Theorems in Sections \ref{NTaP}, \ref{NFT}, \ref{NT2}, and \ref{NT3}  is based on the work of K. A. Brading \cite{Bra2002,Brad2005} and her work with H. R. Brown \cite{BrB2000,BroB2002,BrB2003}.

In Section \ref{NT}, we will review Noether's Theorems in their general form as originally presented. 
We will restrict the application to Lagrangians and the action integral is Section \ref{NTaP} as generally applied in physics and contrast the theorems with Hamilton's Principle.  
We restate the theorems as generally applied in physics with an emphasis on infinitesimal transformations in Sections \ref{NFT}, \ref{NT2} and \ref{NT3}, with examples given in flat space.  In Section \ref{NFT}, Noether's First Theorem is applied to the conservation of energy and a complex scalar field.  In Section \ref{NT2}, Noether's Second Theorem is applied to a complex scalar field,  In Section \ref{NT3}, The Third Theorem is related to superpotentials.         
  In Section \ref{SR} we set up some equations in Special Relativity for comparison with General Relativity.  Of interest are the equations for electric fields and the Stress-Energy Tensor.  
   General Relativity is set up in Section \ref{GRPC} with reference to equations that will be looked at when applying Noether's Theorems.  Besides developing the equations for the electric fields and the Stress-Energy Tensor, the equations for an ideal fluid or gas are derived.
    The application of Noether's First Theorem to General Relativity is examined in Section \ref{GRNT1}.  This application leads to problems with the formulation of the conservation of energy in General Relativity, that was the motivation behind Noether's work of 1918.  Work-arounds use a pseudotensor.  The Landau and Lifshitz pseudotensor and the Dirac pseudotensor are presented as examples.  A general expression for the conservation of energy is possible for maximally symmetric spaces.  This is looked at for the standard model of Cosmology.    
     Noether's Second Theorem and The Third Theorem in the context  of General Relativity  are presented in Section \ref{GRNT2}.
In  Appendix A, variational calculus with variable end points is developed for the one dimensional case and generalised to multiple dimensions.

%%%%%%%%%%%%%%%%%%%%%%%%%%%%%%%%%%%%%%
%%%%%%%%%%%%%%%%%%%%%%%%%%%%%%%%%%%%%%%  
%
\section{Noether's Theorems}
\label{NT}
%
%%%%%%%%%%%%%%%%%%%%%%%%%%%%%%%%%%%%%%%%
In this section, we will review both of Noether's Theorems as originally stated by Noether. In the next section,  we  state them as they are generally stated today when applied to physics.  We begin with following Noether's original work \cite{Noe1918}.  We will confine our remarks to real functions that are continuous and single valued, although the theorems can be applied to complex analytic functions as well.  

We will consider groups of transformations.  Thus each transformation must have an inverse and a composite of two transformations is equivalent to a transformation in the group.  We will call the group $\textgoth{G}_\rho$ if the transformations depend on $\rho$ real parameters.  Whereas the group $\textgoth{G}_{\infty \rho}$ depends on $\rho$ functions and their derivatives.  It is also possible to have mixed groups that depend on functions and on parameters.  

To set the problem up, we consider $n$ independent variables $x_i$ and $\mu$ dependent functions $u_\alpha$.   Then, a  transformation group will transform these to variables $y_j$ and functions $v_\alpha$.  There will be only $n$ independent variables $y$, the others being determined by some of the functions $v$.  This is because the group must contain a member that will transform the $y$'s and $v$'s back to the $x$'s and $u$'s.  A function is an invariant of the group if there is a relationship 

\begin{equation}
P\left(x, u, \frac{\partial u}{\partial x}, ... \right) = P\left(y, v, \frac{\partial v}{\partial y}, ... \right).
\label{ya}
\end{equation}
We will be interested in integrals in particular.  So we will be looking at invariant integrals of the group, 
schematically
\begin{eqnarray}
I =  {\int f\left(x, u, \frac{\partial u}{\partial x}, ... \right) dx} = {\int f\left(y, v, \frac{\partial v}{\partial y}, ... \right) dy},
\label{I1}
\end{eqnarray}
where the integration regions are the same.
One should not infer from these equations that the functions remain form invariant.  We can take the variation of the integral $I$.  Then we form 
\begin{eqnarray}
\delta I =  {\int \delta f dx}.
\label{I2}
\end{eqnarray}
Integrating by parts, we will end up with 
\begin{eqnarray}
\delta I =  {\int \delta f dx} = {\int \left[ \sum \psi_i \left(x, u, \frac{\partial u}{\partial x}, ... \right) \delta u_i \right] dx} + B,~
\label{I3}
\end{eqnarray}
where the $\psi$'s are the functions that we set equal to zero to get the Euler-Lagrange equations, and $B$ is the boundary contributions that we get form the boundary when we integrate by parts and are set equal to zero in Hamilton Principle.  Thus, in the one dimensional case with only the first derivative of $u$, we have the familiar 
\begin{eqnarray}
\psi = \frac{\partial f}{\partial u} 
 - \frac{d}{dx} \frac{\partial f}{\partial (\frac {\partial u}{\partial x})}.
 \label{EL}
\end{eqnarray}
The boundary can be rewritten as an integral over a divergence, so we can write
\begin{eqnarray}
\delta f = \sum \psi_i \left(x, u, \frac{\partial u}{\partial x}, ... \right) \delta u_i + \rm{Div} A,
 \label{f2}
\end{eqnarray}
where
\begin{eqnarray}
\rm{Div} A = \frac{\partial A_i}{\partial x_i}.
 \label{DivA}
\end{eqnarray}
We require that  Div$A$ be linear in the $\delta u$'s which are assumed to be of the same order as for $\delta f$.

For completeness we will state the English translation of Noether's Theorems.  We will not discuss any applications of them at this point but will restate them in the modern interpretation as applied to the cases that we will be discussing.  From the above considerations, Noether proposed two theorems \cite{Tra1971}:
\begin{enumerate}[I.]
\item
If the integral $I$ is invariant with respect to a $\textgoth{G}_{ \rho}$, then $\rho$ linearly independent combinations of the Lagrange expressions become divergences -- and from this, conversely, invariance of $I$ with respect to a $\textgoth{G}_{ \rho}$ will follow.  The theorem holds good even in the limiting case of infinitely many parameters.
\item
If the integral $I$ is invariant with respect to a $\textgoth{G}_{\infty \rho}$ in which the arbitrary functions occur up to the $\sigma$-th derivative, then there subsist $\rho$ identity relationships between the Lagrange expressions and their derivatives up to the $\sigma$-th order.  In  this case also, the converse holds.
\end{enumerate}
%
%
%
%
%%%%%%%%%%%%%%%%%%%%%%%%%%%%%%%%%%%%%%%%%%%%%%%
%%%%%%%%%%%%%%%%%%%%%%%%%%%%%%%%%%%%%%%%%%
%
\section{Noether's Theorem and Physics}
\label{NTaP}
%
%%%%%%%%%%%%%%%%%%%%%%%%%%%%%%%%%%%%%%%%%%%%%%

We will be interested in applying Noether's Theorem to Lagrangians.  We will consider Lagrangians of the form
\begin{eqnarray}
\mathscr{L }=   \mathscr{L}(\phi_{r}, \phi_{r , \nu}, x^{\nu}),
\label{L1}
\end{eqnarray}
where we have labeled the $r$ fields $\phi_r$.  We will be looking at Lagrangians with only the first order derivatives of the fields, however, the equations can be generalised to higher order derivatives.    We form the action 
\begin{eqnarray}
S =  {\int {\mathscr{L} d^4x_\nu}}.
\label{S2}
\end{eqnarray}
Upon taking the variation of $\mathscr{L}$ with respect to $\phi_r$ with fixed boundary and integrating by parts, we have
\begin{eqnarray}
\delta S = {\int \delta{\mathscr{L} d^4x_\nu}} = {\int \sum_r \psi_r \delta \phi_r d^4x_\nu},
\label{II3}
\end{eqnarray}
where 
\begin{eqnarray}
\psi_r \equiv \frac{\delta \mathscr{L}}{\delta \phi_r} =
 \frac{\partial \mathscr{L}}{\partial \phi_r} 
- \frac{\partial}{\partial x^\mu} \left( \frac{\partial \mathscr{L}}{\partial \phi_{r , \mu}} \right).
\label{psi}
\end{eqnarray}
If the Lagrangian contains higher order derivatives of the fields, then there will be added integration by parts and added terms to the right of eq. (\ref{psi}).
The $\psi$'s are often called the Euler Expressions and $\delta \mathscr{L} / \delta \phi$ the Hamiltonian derivative, but they are also called the Lagrangian Expressions, the Lagrangian derivative, the Euler derivative and the functional derivative.  Sometimes the $\psi_r$ will be written as $[\psi]_r$ in order to emphasise that the $\psi$'s are a short hand for the expressions to the far right.  

Hamilton's Principle \cite{Gol1965, LL1V1976}  may be stated as: The first order variation of the action for fixed boundary, that is the variation of the dependent functions (fields), must be zero at the boundary and no variation of the independent variables, must be zero.   
In eq. (\ref{psi}), this would require that $\psi_r = 0$, and gives us the usual Euler-Lagrange equations.    
 
The problem that Noether looked at (NP) should be contrasted with the commonly used Hamilton's Principle (HP)  in physics. 
For Hamilton's Principle, we consider the action $S$ defined by eq. (\ref{S2}).
They differ in four important ways \cite{BrB2003}.  
\begin{enumerate}
\item
NP is not a principle.  HP is a condition that must be met in order to find the equations of motion.  NP asks if certain conditions are met, then one can ascertain certain results.
\item
For HP the variation is considered arbitrary.  For NP the variation is specific.
\item
The variations considered in NP can involve the dependent variables and the independent variables.  The variation considered in HP only consider the dependent variable.
\item
In NP the variations are not required to be zero at the boundary, whereas in HP this is always the case.
\end{enumerate}

We will develop NT for our Lagrangians for infinitesimal transformations.   
This will be following the development in Brading and Brown \cite{BrB2003}.  
Consider some infinitesimal transformation.  
We start by looking at the change in the action for the transformation,  
\begin{eqnarray}
\Delta S =
\nonumber \bar{{S}}(\bar{\phi}_ s(\bar{x} ^\nu), \bar{\phi} _{s , \mu}(\bar{x}^\nu), \bar{x}^\nu )~~~ \\
-{S}({\phi}_ s({x} ^\nu), {\phi} _{s , \mu}({x}^\nu), {x}^\nu ) ,
\label{DS}
\end{eqnarray}
or 
\begin{eqnarray}
\nonumber \Delta S = {\int _{\bar{\Omega}} {\bar {\mathscr{L}}(\bar{\phi}_ s(\bar{x} ^\nu), \bar{\phi} _{s , \mu}(\bar{x}^\nu), \bar{x}^\nu )  d^4\bar{x}_\nu}}~~~ \\
-{\int_\Omega {\mathscr{L}(\phi_ s(x ^\nu), \phi _{s , \mu}(x^\nu), x^\nu )  d^4x_\nu}}.
\label{DS2}
\end{eqnarray}
We may proceed in two different ways.  First, is to require that the transformed $\bar{S}$ satisfies HP if $S$ does.  In this case, the difference of the action do to the transformation will be equal to a divergence,
\begin{eqnarray}
\Delta S =  {\int_\Omega { \partial_\mu \Lambda^\mu d^4x_\nu}}.
\label{DS3}
\end{eqnarray}
This is because the Euler Expressions of a total divergence vanishes \cite{Noe1918}.  This condition will be met if the Lagrangian is form invariant, so we have 
\begin{eqnarray}
\bar{\mathscr{L}}({\phi}_ s({x} ^\nu), {\phi} _{s , \mu}({x}^\nu), {x}^\nu ) 
= \mathscr{L}({\phi}_ s({x} ^\nu), {\phi} _{s , \mu}({x}^\nu), {x}^\nu ) .
\label{L3}
\end{eqnarray}
Thus, the first treatment is to require that the Lagrangian is strictly form invariant \cite{BrB2003}.   

The second way is to treat the action as numerically invariant.  This would require that $\Delta S$ in eq. (\ref{DS2}) be zero.  However, this would allow the Lagrangian $\bar{\mathscr{L}}$ to differ from $ \mathscr{L}$ by a divergence term \cite{BrB2003},
\begin{eqnarray}
\nonumber \bar{\mathscr{L}}({\phi}_ s({x} ^\nu), {\phi} _{s , \mu}({x}^\nu), {x}^\nu ) 
=&& \mathscr{L}({\phi}_ s({x} ^\nu), {\phi} _{s , \mu}({x}^\nu), {x}^\nu ) \\
&&+ \partial_\mu \Theta^\mu.
\label{L4}
\end{eqnarray}
In this case, we have 
\begin{eqnarray}
\bar{\mathscr{L}} - {\mathscr{L}} =    \partial_\mu \Theta^\mu .
\label{DS13}
\end{eqnarray}
So, in both cases, we find that the change in the action is an integral over a total divergence.   This agrees with Noether's original paper, as it must.

In the appendix, the variation of the end points is worked out.  There, we considered the Lagrangian to be form invariant.  For a general transformation, this leads to the most general form of NT for the  Lagrangian given in eq. (\ref{L1}) as 
%%
%%
%** \begin{widetext}
%%
%%
%
%
\begin{eqnarray}
 \Delta S = 0 =  \int_\Omega  \bigg\{
\psi_s \delta \phi_s 
 +  \bigg{[}  \bigg{(}  
     \mathscr{L} \delta^\mu_\gamma
  - \frac{\partial \mathscr{L}}{\partial  \phi_{s,\mu}} { \phi_{s, \gamma}} 
 \bigg{)}
 \delta x^\gamma \nonumber \\
  +\frac{\partial \mathscr{L}}{\partial \phi_{ s,\mu}} \delta \phi_s 
 \bigg{]}_{, \mu}
  \bigg\} d^4 x_\nu,~~~
\label{nLg16}
\end{eqnarray}
where the second term in the integral is in the form of a divergence.  
If the transformed Lagrangian satisfies HP, then $\psi_s=0$.  Many of the transformations that we will be examining satisfy HP.  However, one should not assume that a transformation will satisfy HP.

We can extract from eq. (\ref{nLg16}) the relationship
\\
\begin{eqnarray}
\sum_s \psi_s \delta \phi_s = ~~~~~~~~~~~~~~~~~~~~~~~~~~~~~~~~~~~~~~~~~~~~~~ \nonumber \\
    - \sum_s \left\{ 
  \bigg{[}  
     \mathscr{L} \delta^\mu_\gamma - \frac{\partial \mathscr{L}}{\partial  \phi_{s,\mu}} { \phi_{s, \gamma}} 
 \bigg{]}
 \delta x^\gamma 
  + \frac{\partial \mathscr{L}}{\partial \phi_{ s,\mu}} \delta \phi_s 
 \right\}_{, \mu}.
\label{nLg17}
\end{eqnarray}
%
%
%%
%%
%** \end{widetext}
%% 
%%
This can be written in a more general form as \cite{BrB2000,Bra2002} 
\begin{eqnarray}
\sum_s \psi_s \delta \phi_s = 
   -  \sum_s B^\mu_{s,\mu},
\label{nLg18}
\end{eqnarray}
where
\begin{eqnarray}
B^\mu_s=    \bigg{[}  
     \mathscr{L} \delta^\mu_\gamma
  - \frac{\partial \mathscr{L}}{\partial  \phi_{s,\mu}} { \phi_{s, \gamma}} 
 \bigg{]}
 \delta x^\gamma 
  +\frac{\partial \mathscr{L}}{\partial \phi_{ s,\mu}} \delta \phi_s 
 - \Delta \Gamma^\mu,~
\label{nLg19}
\end{eqnarray}
where, in general, $B^\mu$ can include the term $\Delta \Gamma^\mu$ where $\Gamma^\mu$ is an arbitrary function that is linear in the $\delta \phi$'s.  For our infinitesimal transformation, this term is infinitesimal and  can account for any ambiguity in the transformation or in the Lagrangian.  Quite often we set the $\Gamma$ term equal to zero.
We will consider eq. (\ref{nLg18}) to be the most general form of NT. This equation agrees with Noether's eq. (12).    

A notational note:  It will be remembered that the first term on the RHS above, is to account for the variation of the fields at the boundary caused by varying the boundary (this is clearly shown for the  1-D case in the appendix).  Since this can be considered a variation of the fields, this term can be include it in a total variation of the fields.  This is usually done in two ways.  One, by  defining our $\delta \phi$ as $\delta_0 \phi$, the direct variation of the fields.  Then the total variation of the fields is given by
\begin{eqnarray}
\delta \phi_s = \delta_0 \phi_s -  \phi_{s,\gamma} \delta x^\gamma.
\label{dphi1}
\end{eqnarray}
Then eq. (\ref{nLg17}) becomes
\begin{eqnarray}
\sum_s \psi_s \delta_0 \phi_s = -
  \sum_s  \left\{ 
    \mathscr{L} 
 \delta x^\mu 
  +\frac{\partial \mathscr{L}}{\partial \phi_{ s,\mu}} \delta \phi_s 
 \right\}_{, \mu}.
\label{nLg20}
\end{eqnarray}
Sometimes this total variation of the fields is denoted as $\bar{\delta} \phi_s$.

%
%%%%%%%%%%%%%%%%%%%%%%%%%%%%%%%%%%%%%%%%%%%%%%%%%%%%%%%%%%%%%%%%%%%%%%%%%%%%%%%%%%%%%%%%%%
%
%
\section{ Noether's First Theorem}
\label{NFT}
%
%
%%%%%%%%%%%%%%%%%%%%%%%%%%%%%%%%%%%%%%%%%%%%%%%%%%
%
In this section, we will examine the use of Noether's first theorem in physics.  We will start by rewriting our general form as to make the difference between the infinite transformation and our original variation more transparent.  We will then give a statement of the theorem and some examples.  We start with our definition of the single parameter variation of a quantity $X$ \cite{Gol1965n1},

%
%%%%%%%%%%%%%%%%%
%
%
%
%
\begin{eqnarray}
\delta X = \left(\frac{\partial X}{\partial \alpha}\right)_{\alpha = 0} d\alpha,
\label{Xcon}
\end{eqnarray}
where $\alpha$ is the parameter.  Here, X is  any variable that is to be varied.  We rewrite this as 
\begin{eqnarray}
\delta X \Rightarrow \left(\frac{\partial  X}{\partial  \omega_i}\right) \Delta \omega_i,
\label{Xcon2}
\end{eqnarray}
for the $i$th parameter $\omega_i$ of the transformation.  Following Brading's work,  Noether's first theorem, for our action and an infinitesimal transformation, can be stated as \cite{Bra2002}:
\begin{description}
\item[\textbf{Theorem 1.}]
\textit{ If the action S is invariant under a continuous group of transformations depending smoothly on $\rho$ independent constant parameters $\omega_i (i = 1, 2,   \rho)$, then there are $\rho$ relationships} 
\begin{eqnarray}
\sum_s \psi_s \frac{\partial  \phi_s}{\partial  \omega_i} = 
   \partial_\mu j^\mu_i,
\label{nNT1}
\end{eqnarray}
\textit {where $j^\mu_i$ is the Noether current associated with the parameter $\omega_i$.}  
\end{description}
Equation (\ref{nNT1}) agrees with Noether's eq. (13).  This type of symmetry is called global or proper.
We can derive this expression from eq. (\ref{nLg18}) with $B^\mu_s$ given by  eq. (\ref{nLg19}), by substituting in eq. (\ref{Xcon2}), 
%%
%%
%***\begin{widetext}
%%
%%
%
%
\begin{eqnarray}
\sum_s \psi_s \frac{\partial  \phi_s}{\partial  \omega_i} \Delta \omega_i =
    - \sum_s \Bigg\{ 
  \bigg{[}  
   \mathscr{L} \delta^\mu_\gamma
    - \frac{\partial \mathscr{L}}{\partial  \phi_{s,\mu}} { \phi_{s, \gamma}} 
 \bigg{]}
  \frac{\partial  x^\gamma}{\partial  \omega_i} \Delta \omega_i \nonumber \\
  +\frac{\partial \mathscr{L}}{\partial \phi_{ s,\mu}} \frac{\partial  \phi_s}{\partial  \omega_i} \Delta \omega_i
 -\frac{\partial  \Gamma ^\mu}{\partial  \omega_i} \Delta \omega_i  \Bigg\}_{, \mu}~~~~~~
\label{nNT2}
\end{eqnarray}
Since the parameters $\Delta \omega_i$ are constants, i.e. do not depend on the coordinates, they can be taken outside the bracket and cancelled from both sides, giving eq. (\ref{nNT1}), with 
\begin{eqnarray}
j^\mu_i = 
    - \sum_s 
  \Bigg\{ \bigg{[}  
   \mathscr{L} \delta^\mu_\gamma
 -  \frac{\partial \mathscr{L}}{\partial  \phi_{s,\mu}} { \phi_{s, \gamma}} 
 \bigg{]}
  \frac{\partial  x^\gamma}{\partial  \omega_i} \nonumber \\
  + \frac{\partial \mathscr{L}}{\partial \phi_{ s,\mu}} \frac{\partial  \phi_s}{\partial  \omega_i}
  - \frac{\partial  \Gamma^\mu}{\partial  \omega_i} \Bigg\}.
\label{nNT3}
\end{eqnarray}
%
%
%%
%%
%***\end{widetext}
%%
%%
This equation is the general expression for Noether currents, as such we have included the $\Gamma^\mu$ term.  If the transformation satisfies HP, then the $\psi$'s can be set to zero in eq. (\ref{nNT1}), and the Noether currents are conserved.  We have 
\begin{eqnarray}
\partial_\mu j^\mu_i = 0.
\label{Nj1}
\end{eqnarray}

Note that both eqs. (\ref{nNT3}) and (\ref{Nj1}) hold for each value of $i$.  Equation (\ref{Nj1}) is the continuity equation for the Noether current $j^\mu_i$.  Integration of the equation for suitable boundary conditions will give the Noether charge,  
\begin{eqnarray}
Q_i = \int_{V}  {j^0_i d^3 x}.
\label{nNT7}
\end{eqnarray}
There will be a conserved charge for each of the $i$ conserved currents.
Below we give two examples.  
%%%%%%%%%%%%%%%%%%
%
%
\subsection{Stress-Energy Tensor }
%
%
%%%%%%%%%%%%%%%%%%%
%
We start by finding the  stress-energy tensor by considering an infinitesimal coordinate transformation,   
for each coordinate independently,
\begin{eqnarray}
\frac{\partial  x^\gamma}{\partial  \omega_i} = \delta_i^\gamma,
\label{dx11}
\end{eqnarray}
where $i$ runs over the four coordinates.    
These four infinitesimal transformations will leave $\delta S = 0$.  
 For some Lagrangian,  each $i$ will give a Noether current. 
For this case, we can apply Noether's First Theorem and we will have
\begin{eqnarray}
j^\mu_i =   \sum_s  
  \bigg{[}  
   \mathscr{L} \delta^\mu_\gamma
 - \frac{\partial \mathscr{L}}{\partial  \eta_{s,\mu}} { \eta_{s, \gamma}} 
 \bigg{]}
 \delta^\gamma_i,   
\label{dx12}
\end{eqnarray}
which we write for all four currents as the stress-energy tensor
\begin{eqnarray}
 j^\mu_i = T_i^\mu = - \sum_s  \frac{ \partial \mathscr {L}}{{\partial (\eta _{s ,\mu})}} \eta_{s ,i} 
 + \delta_i^\mu \mathscr{L}.
\label{dx13}
\end{eqnarray}
Since the Euler expressions for this transformation are zero,  we have four conserved currents, one for each dimension,  
\begin{eqnarray}
  j_{i ,\mu}^\mu = T_{i ,\mu}^\mu = 0.
\label{Tik3}
\end{eqnarray}
Note that we used the term stress-energy tensor.  Some authors will call this the energy-momentum tensor.

Integrating over an appropriate volume, we have the continuity equation for the $i=0$ current,  
\begin{eqnarray}
P^\mu = \int_{V}  {j^{ 0 \mu} d^3 x} = \int_{V}  {T^{  0 \mu} d^3 x} ,
\label{4dp}
\end{eqnarray}
with
\begin{eqnarray}
P_{~,\mu}^\mu  = 0,
\label{4dp2}
\end{eqnarray}
where $P^{\mu}$ is the momentum 4-vector. We identify the $0$ component as the energy.  The other three components are the three components of the 3-momentum.  The continuity equation tells us that the \textquotedblleft charge\textquotedblright (energy) is conserved and the time rate of change is equal to the flux of charge out of the system (divergence of the momentum = change of energy).

Likewise, the other currents can be found for $i = 1, 2~ \& ~3$.  For these cases, the Noether charges are the three values of the momentum density.  The three spacial components of the continuity equation are  the $\mu = 1, 2~ \& ~3$ components of the flux of momentum density change in the system for each momentum density component.  The nine values    $i = 1, 2~ \& ~3$ with   $\mu = 1, 2~ \& ~3$, make up the components of the stress tensor.  
%
%
%%%%%%%%%%%%%%%%%%
%
%
%
%%%%%%%%%%%%%%%%%%
%
%
\subsection{Complex Scalar Field}
%
%
%%%%%%%%%%%%%%%%%%%
%
Another example of Noether's First Theorem can be found for a scalar field.  
We start with the Lagrangian for complex fields \cite{LL4V1971},
\begin{eqnarray}
\mathscr{L} = \partial_\nu \phi^* \partial^\nu \phi - m^2 \phi^* \phi
 = g^{\nu \mu} \partial_\nu \phi^* \partial_\mu \phi - m^2 \phi^* \phi .
\label{I42}
\end{eqnarray}
 This Lagrangian is invariant under transformations of the type
\begin{eqnarray}
 \phi \rightarrow e^{i \alpha} \phi,~~~~~~~~  \phi^* \rightarrow e^{-i \alpha} \phi^*,
\label{phi2}
\end{eqnarray}
where $\alpha$ is a constant, i.e. not a function of $x$.  
This type of transformation is called a gauge transformation of the first kind.  Since the Lagrangian is invariant, the Euler-Lagrange equations will still be satisfied and $\psi_i = 0$.   For an infinitesimal transformation, we can expand the exponential and find, to lowest order
\begin{eqnarray}
 \phi \rightarrow \phi + {i \alpha} \phi,~~~~~~~~  \phi^* \rightarrow \phi^*  {-i \alpha} \phi^*.
\label{phi22}
\end{eqnarray}
For this transformation we have,
\begin{eqnarray}
\frac{\partial  \phi}{\partial   \alpha}= i \phi, ~~~~~~~~ 
\frac{\partial  \phi^*}{\partial  \alpha}= i \phi^*,
\label{N22}
\end{eqnarray}
and
\begin{eqnarray}
\frac{\partial  x^\gamma}{\partial  \alpha}= 0.
\label{sF133}
\end{eqnarray}
 Subbing into eq. (\ref{nNT2})  from eqs. (\ref{N22}) and (\ref{sF133}), to find
\begin{eqnarray}
j^\nu_{, \nu} = 0,
\label{juu}
\end{eqnarray}
with
\begin{eqnarray}
j^\nu = i (\phi^* \partial^\nu \phi - \phi ~ \partial^\nu \phi^* ).
\label{ju}
\end{eqnarray}
This is a continuity equation.  The $j^0$ component integrated over a volume is the charge, with the spacial components   representing the flow of charge at the surface, similar to what was found for the momentum vector above.  

%%%%%%%%%%%%%%%%%%%%%%%%%%%%%%%%%%%%%%%%%%%
%%%%%%%%%%%%%%%%%%%%%%%%%%%%%%%%%%%%%%%%%%%%
%
%
\section{ Noether's Second Theorem}
\label{NT2}
%
%
%%%%%%%%%%%%%%%%%%%%%%%%%%%%%%%%%%%%%%%%%%%%%
%
%
We consider transformations of the fields by $\rho$ arbitrary functions. We can choose the function that are zero at the boundary.  This will limit our application to the inclosed area.  We now state Noether's Second Theorem, for our action and an infinitesimal transformation $\Delta p_i$, from Brading \cite{Bra2002}:
\begin{description}
\item[\textbf{Theorem 2.}]
\textit{ If the action S is invariant under a continuous group of transformations depending smoothly on $\rho$ independent arbitrary functions $p_i(x)$ and their first derivatives, then there are $\rho$ relationships} 
\begin{eqnarray}
\sum_s \psi_s a_{is} = 
\sum_s   \partial_\mu (\psi_s b^\mu_{is}),
\label{nNT20}
\end{eqnarray}
\textit {where $a_{is}$ and $b^\mu_{is}$ are functions of $\phi, \partial_\mu \phi$ and $x^\mu$.}  
\end{description}
We write this  theorem with the restriction of first partial derivatives for the fields and for the transformation.  This restriction can be generalised to higher order derivatives of both \cite{Noe1918}.

We can derive this expression from eq. (\ref{nLg16}) by substituting in eq. (\ref{nLg19}), 
\begin{eqnarray}
\Delta S =  \int_\Omega {\sum_s \left(
\psi_s \delta \phi_s + 
B^\mu_{s, \mu}
  \right) d^4 x_\nu} = 0.
\label{nNT21}
\end{eqnarray}
We require the transformation to vanish at the boundary, thus the second term gives zero, so
\begin{eqnarray}
\Delta S =  \int_\Omega {\sum_s \left(
\psi_s \delta \phi_s 
  \right) d^4 x_\nu} = 0.
\label{nNT22}
\end{eqnarray}
We can write out an expression for $\delta \phi_s$ for each of the transformation functions as
\begin{eqnarray}
\delta \phi_s =&& \frac {\partial \phi_s}{\partial  p_i} \Delta p_i 
+ \frac {\partial \phi_s}{\partial (\partial_\mu p_i)} \partial_\mu \Delta p_i \nonumber \\
=&& a_{is} \Delta p_i + b^\mu_{is} \partial_\mu \Delta p_i .
\label{nNT23}
\end{eqnarray}
Subbing this into eq. (\ref{nNT22}) we have
\begin{eqnarray}
\Delta S =  \int_\Omega {\sum_s  \psi_s \left(
a_{is} \Delta p_i + b^\mu_{is} \partial_\mu \Delta p_i  
  \right) d^4 x_\nu} = 0.
\label{nNT24}
\end{eqnarray}
The second term can be integrated by parts to give 
\begin{eqnarray}
\Delta S =  \int_\Omega {\sum_s  \left[\psi_s 
a_{is} - \partial_\mu (\psi_s b^\mu_{is})   
  \right]  \Delta p_i  d^4 x_\nu} = 0.
\label{nNT25}
\end{eqnarray}
Equating the bracket to zero gives Noether's Second Theorem.  
We see that Noether's Second Theorem is valid inside the boundary.  This is called an improper transformation and leads to local symmetry groups.  
%
%%%%%%%%%%%%%%%%%%%%%%%%%%%%%%%%%%%%%
%%%%%%%%%%%%%%%%%%%%%%%%%%%%%%%%%%%%%%
%
%
%%%%%%%%%%%%%%%%%%
%
%
\subsection{Complex Scalar Field }
%
%
%%%%%%%%%%%%%%%%%%%
%
In this section we re-examine the complex scalar field by identifying the charge as electric charge and adding an external electric-magnetic field, although the identification is not necessary and the charge and fields can have other associations.  We follow Ref. \cite{BrB2000} and start with the Lagrangian
\begin{eqnarray}
\mathscr{L} = \mathcal{D}_\nu \phi^* \mathcal{D}^\nu \phi - m^2 \phi^* \phi -\frac{1}{4}F^{\mu\nu}F_{\mu\nu}~~~~~ \nonumber \\
 = g^{\nu \mu} \mathcal{D}_\nu \phi^* \mathcal{D}_\mu \phi - m^2 \phi^* \phi -\frac{1}{4}F^{\mu\nu}F_{\mu\nu},
\label{sF21}
\end{eqnarray}
where $\mathcal{D}_\nu = \partial_\nu + i q A_\nu$,  often called the covariant derivative, and $q$ is the charge of the field.
 We consider transformations of  this Lagrangian of the type
\begin{eqnarray}
 \phi \rightarrow e^{i q \theta} \phi,~~~  \phi^* \rightarrow e^{-i q \theta} \phi^*, ~~~
 A_\nu \rightarrow A_\nu + \partial_\nu \theta,
\label{sF22}
\end{eqnarray}
where $\theta = \theta(x)$.
This type of transformation is called a gauge transformation of the second kind.  For an infinitesimal transformation, we can expand the exponential and find, to lowest order,
\begin{eqnarray}
 \phi \rightarrow \phi + {i q \Delta\theta} \phi,~~~\phi^* \rightarrow \phi^*  {-i q \Delta \theta} \phi^*, \nonumber \\
 A_\nu \rightarrow A_\nu + \partial_\nu \Delta \theta.
\label{sF23}
\end{eqnarray}
This gives us the following,
\begin{eqnarray}
 \delta \phi = {i q \Delta\theta} \phi,~~~  \delta \phi^* =  {-i q \Delta \theta} \phi^*, ~~~
 \delta A_\nu = \partial_\nu \Delta \theta.
\label{sF23b}
\end{eqnarray}

Calling $\phi$ our \textquotedblleft1\textquotedblright~field,  $\phi^*$ our \textquotedblleft2\textquotedblright~field and $A_\mu$ our \textquotedblleft3\textquotedblright~field, and comparing to eq. (\ref{nNT23}) we have,
\begin{eqnarray}
a_1 = i q \phi, ~~~~~~~~ a_2 = -i q \phi^*, ~~~~~~~~~~ a_3 = 0,
\label{sF24}
\end{eqnarray}
and 
\begin{eqnarray}
b^\mu_1 = 0, ~~~~~~~~ b^\mu_2 = 0, ~~~~~~~~~~ b^\mu_3 = \delta^\mu_\nu.
\label{sF24b}
\end{eqnarray}
Subbing into eq. (\ref{nNT20}) yields,
\begin{eqnarray}
 &&~~\left[\frac{\partial \mathscr{L}}{\partial \phi} 
- \frac{\partial}{\partial x^\mu} \left( \frac{\partial \mathscr{L}}{\partial \phi_{ ,\mu}} \right)\right]
(iq\phi) \nonumber \\
 &&+\left[\frac{\partial \mathscr{L}}{\partial \phi^*} 
- \frac{\partial}{\partial x^\mu} \left( \frac{\partial \mathscr{L}}{\partial \phi^*_{ ,\mu}} \right)\right]
(-iq\phi^*)  \nonumber \\
 &&= \partial_\mu \left[\frac{\partial \mathscr{L}}{\partial A_\mu} 
- \frac{\partial}{\partial x^\gamma} \left( \frac{\partial \mathscr{L}}{\partial A_{\mu ,\gamma}} \right)\right],
\label{NT2cf}
\end{eqnarray}
from which we find for our Lagrangian
\begin{eqnarray}
\partial_\mu \partial_\nu F^{\mu\nu} = 0.
\label{NT2cf2}
\end{eqnarray}

We can also find the conserved current for this Lagrangian by varying the fields $A^\mu$.  We will have 
\begin{eqnarray}
 \frac{\partial \mathscr{L}}{\partial A_\mu} 
- \frac{\partial}{\partial x^\nu} \left( \frac{\partial \mathscr{L}}{\partial A_{\mu ,\nu}} \right) = 0.
\label{NT2A1}
\end{eqnarray}
Using our Lagrangian eq. (\ref{sF21}) we find 
\begin{eqnarray}
 \partial_\nu F^{\mu\nu} = - j^\mu,
\label{NT2cf3}
\end{eqnarray}
where 
\begin{eqnarray}
j^\mu = i q (\phi^* \mathcal{D}^\mu \phi - \phi ~ \mathcal{D}^\mu \phi^* ).
\label{sF26}
\end{eqnarray}
From eq. (\ref{NT2cf2}) we find
\begin{eqnarray}
 \partial_\mu \partial_\nu F^{\mu\nu} = 0 = \partial_\mu  j^\mu,
\label{NT2cf4}
\end{eqnarray}
and the current is conserved.  This result could be found directly by setting the RHS in eq. (\ref{NT2cf}) equal to zero and subbing in for the Lagrangian, giving,
\begin{eqnarray}
j^\mu_{, \mu} = 0.
\label{NT2cf5}
\end{eqnarray}
%
%

%%%%%%%%%%%%%%%%%%%%%%%%%%%%%%%%%%%%%%%%%%%
%%%%%%%%%%%%%%%%%%%%%%%%%%%%%%%%%%%%%%%%%%%%
%
%
\section{ The Third Theorem}
\label{NT3}
%
%
%%%%%%%%%%%%%%%%%%%%%%%%%%%%%%%%%%%%%%%%%%%%%
%
%
Although Noether gave only two theorems, there is a set of equations that can be derived for the boundary, based on the Second Theorem.  This set of equations are not independent from the second theorem, however, they give a different way to deal with the boundary and are called The Third Theorem, The Boundary Theorem or sometimes  Noether's Third Theorem.  

Once again we consider transformations of the fields by $\rho$ arbitrary functions. We can choose the functions that are zero at the boundary.  This will limit our application to the inclosed area.  Remember, that we are limiting our $\phi_s$ and $p_i$'s to first order derivatives with respect to $x^\mu$.  We now state a third theorem, for our action and an infinitesimal transformation $\Delta p_i$ \cite{BrB2000, BrB2003}:
\begin{description}
\item[\textbf{Theorem 3.}]
\textit{ If the action S is invariant under a continuous group of transformations depending smoothly on $\rho$ independent arbitrary functions $p_i(x)$ and their first derivatives, then there are $\rho$ relationships} 
\begin{subequations}
\label{nNT31}
\begin{eqnarray}
\sum_s \psi_s a_{is} = - \sum_s
   \partial_\mu \left(\frac{\partial \mathscr{L}}{\partial(\partial_\mu \phi_s)} a_{is}\right) \label {nNT31a}
   \\
\sum_s   \partial_\mu (\psi_s b^\mu_{is}) = 
    \partial_\mu j^\mu_i~~~~~~~~~~~~~~  \label {nNT31b}
\end{eqnarray}
\end{subequations}
\begin{eqnarray}
\sum_s \psi_s b^\mu_{is} = - \sum_s
  \left[ \frac{\partial \mathscr{L}}{\partial(\partial_\mu \phi_s)} a_{is}
  + \partial_\nu \left(\frac{\partial \mathscr{L}}{\partial(\partial_\nu \phi_s)} b^\mu_{is}\right) 
  \right],
\label{nNT32}
\end{eqnarray}
\begin{eqnarray}
0 = \sum_s
  \left[ \frac{\partial \mathscr{L}}{\partial(\partial_\mu \phi_s)} b^\nu_{is}
  + \frac{\partial \mathscr{L}}{\partial(\partial_\nu \phi_s)} b^\mu_{is} \right],
\label{nNT33}
\end{eqnarray}
\textit {where $a_{is}$ and $b^\mu_{is}$ are functions of $\phi, \partial_\mu \phi$ and $x^\mu$.}  
\end{description}
To derive these equations, we start with eq. (\ref{nLg17}), with $\delta x^\gamma = 0$,
\begin{eqnarray}
\sum_s \psi_s \delta \phi_s = 
    - \sum_s \left\{ 
  \frac{\partial \mathscr{L}}{\partial \phi_{ s,\mu}} \delta \phi_s 
 \right\}_{, \mu}.
\label{nLgnNT3}
\end{eqnarray}
We sub in for the $\delta \phi_s$  from eq. (\ref{nNT23}),
\begin{eqnarray}
\delta \phi_s  
= a_{is} \Delta p_i + b^\mu_{is} \partial_\mu \Delta p_i ,
\label{nNT3p2}
\end{eqnarray}
and find
\begin{eqnarray}
&&\sum_s \psi_s (a_{is} \Delta p_i + b^\mu_{is} \partial_\mu \Delta p_i ) \nonumber \\
=     &&- \sum_s \partial_\mu\left\{ 
  \frac{\partial \mathscr{L}}{\partial \phi_{ s,\mu}} 
 ( a_{is} \Delta p_i + b^\nu_{is} \partial_\nu \Delta p_i  )\right\}.
\label{nLgnNTp3}
\end{eqnarray}
Expanding out the partial derivative on the RHS gives
\begin{eqnarray}
\sum_s \psi_s (a_{is} \Delta p_i + b^\mu_{is} \partial_\mu \Delta p_i ) = -  \sum_s \Big\{ 
  \partial_\mu \Big( \frac{\partial \mathscr{L}}{\partial \phi_{ s,\mu}} a_{is} \Big) \Delta p_i   \nonumber \\ 
 +  \frac{\partial \mathscr{L}}{\partial \phi_{ s,\mu}} a_{is} \partial_\mu \Delta p_i  
+ \partial_\mu \Big(\frac{\partial \mathscr{L}}{\partial \phi_{ s,\mu}} b^\nu_{is}\Big) \partial_\nu \Delta p_i  \nonumber \\
+\frac{\partial \mathscr{L}}{\partial \phi_{ s,\mu}} b^\nu_{is} \partial_{\mu \nu} \Delta p_i  \Big\}.~~~~~~~~~~~~~~~~~~~~~~~~~
\label{nLgnNTp4}
\end{eqnarray}
Equating coefficients of $\Delta p_i, \partial_\mu \Delta p_i,$ and $\partial_{\mu\nu} \Delta p_i$ leads to the eqs. (\ref{nNT31a}), (\ref{nNT32}) and (\ref{nNT33}).  We can find eq. (\ref{nNT31b}) from eq. (\ref{nNT31a}) by using Noether's Theorem 2 for the LHS and using eq. (\ref{nNT3}) for a Noether current.  We find
\begin{eqnarray}
j^\mu_i = 
    - \sum_s \left\{ 
 \frac{\partial \mathscr{L}}{\partial \phi_{ s,\mu}} \frac{\partial  \phi_s}{\partial  \omega_i} \right\}
=     - \sum_s \left\{ 
 \frac{\partial \mathscr{L}}{\partial \phi_{ s,\mu}} a_{is} \right\},
\label{nNT3p5}
\end{eqnarray}
and
\begin{eqnarray}
\sum_s \psi_s a_{is} &&= \sum_s   \partial_\mu (\psi_s b^\mu_{is})  \nonumber \\ &&= - \sum_s
   \partial_\mu \left(\frac{\partial \mathscr{L}}{\partial(\partial_\mu \phi_s)} a_{is}\right) 
   + \partial_\mu j^\mu_i.
\label{nNTnew5}
\end{eqnarray}
%
%
%
%
%
%%%%%%%%%%%%%%%%%%%%%%%%%%%%%%%%%%%%%
%%%%%%%%%%%%%%%%%%%%%%%%%%%%%%%%%%%%%%
%
%
%
%
%
\subsection{\textquotedblleft Superpotentials\textquotedblright }
%
%
%%%%%%%%%%%%%%%%%%%
%
We include this section for completeness.  
Staring with eq. (\ref{nNT31b}), we can rewrite it as
\begin{eqnarray}
 \Theta^\mu_i  = 
  j^\mu_i  - \sum_s   \psi_s b^\mu_{is}
\label{SP1}
\end{eqnarray}
with
\begin{eqnarray}
 \partial_\mu \Theta^\mu_i   = 0.
\label{SP2}
\end{eqnarray}
From this we can define superpotentials $U^{\mu \nu}_i$ by \cite{Brad2005} 
\begin{eqnarray}
 \Theta^\mu_i   =  \partial_\nu U^{\mu \nu}_i,
\label{SP3}
\end{eqnarray}
and we have
\begin{eqnarray}
  \partial_\mu \partial_\nu U^{\mu \nu}_i = 0.
\label{SP4}
\end{eqnarray}
Then we can write a more general Noether current as 
\begin{eqnarray} 
  j^\mu_i  = \sum_s   \psi_s b^\mu_{is} + \partial_\nu U^{\mu\nu}_i.
\label{SP5}
\end{eqnarray}

This will give us a more general expression for our Noether current and allow us more flexibility in modelling unknown components to Noether currents.
%
%%%%%%%%%%%%%%%%%%%%%%%%%%%%%%%%%%%%%%%%%%%
%%%%%%%%%%%%%%%%%%%%%%%%%%%%%%%%%%%%%%%%%%%
%
%%%%%%%%%%%%%%%%%%%%%%%%%%%%%%%%%%%%%%%%%%%%
%
%
\section{Special Relativity Review}
\label{SR}
%
%
%%%%%%%%%%%%%%%%%%%%%%%%%%%%%%%%%%%%%%%%%%%%%
%
%
%
We review some equations in Special Relativity (SR) to set up corresponding equation in General Relativity (GR) for comparison.

%%%%%%%%%%%%%%%%%%%%%%%%%%%%%%%%%%%%%
%%%%%%%%%%%%%%%%%%%%%%%%%%%%%%%%%%%%%%
%
%
%
%
%
\subsection{Equation of Motion }
%
%
%%%%%%%%%%%%%%%%%%%
%
In mechanics we find that an object will travel in a straight line if no force acts on it.  For gravity, we can solve HP to find Newtons Law
\begin{eqnarray} 
 m \frac {d^2 x^i}{dt^2} = \mathbf{F}^i
   = - \mathbf{\nabla} V_g.
\label{SRa1}
\end{eqnarray}
where the force is given by the negative gradient of the gravitational potential.  In the absence of a potential, the equation of motion is a straight line.  

Although this works fine for material particles, the propagation of light poses problems, since it travels along a null vector, it has an elapsed proper time of zero.  To get equations of motion, we need to parameterise its path with something other than proper time.  

Looking at the propagation vector $\mathbf{k}$ for light, we generalise this to 4-D as $k^\mu = (\omega, \mathbf{k})$, with $c = 1$.  This gives the condition for the propagation of light as \cite{LL2V1975} 
\begin{eqnarray} 
 k^\mu k_\mu = 0.
\label{SRa2}
\end{eqnarray}
We can define the eikonal, $\Psi$ as \cite{LL2V1975n2}  
\begin{eqnarray} 
 k_\mu = -\partial _\mu \Psi.
\label{SRa3}
\end{eqnarray}
This leads to the eikonal equation 
\begin{eqnarray} 
\eta^{\mu\nu}  \partial _\mu \Psi \partial_\nu \Psi = 0,
\label{SRa4}
\end{eqnarray}
where $\eta^{\mu\nu}$ is the usual Minkowskian space metric with $\eta^{00} = + 1$

%
%%%%%%%%%%%%%%%%%%%%%%%%%%%%%%%%%%%%%
%%%%%%%%%%%%%%%%%%%%%%%%%%%%%%%%%%%%%%
%
%
%
%
%
\subsection{Electric Fields}
%
%
%%%%%%%%%%%%%%%%%%%
%
Maxwell equations can be written in differential form as \cite{LL2V1975n1} % LL eq. (26.1, 26.2, 30.3, 30.4); D, eq. (23.3 - 23.6)
\begin{subequations}
\label{Maxd1}
\begin{eqnarray}
\mathbf{\nabla \cdot B} =  0  \label {Maxd1a}
   \\
\mathbf{ \nabla \times E} +  \frac{\partial \mathbf{B}}{\partial t}
 =   0  \label {Maxd1b}
   \\
\mathbf{\nabla \cdot E} = 4 \pi  \rho  \label {Maxd1c}
   \\
\mathbf{ \nabla \times B} - \frac{\partial \mathbf{E}}{\partial t}
 =  4 \pi  \mathbf{J}  \label {Maxd1d}  
\end{eqnarray}
\end{subequations}
In the usual way, we introduce the four vector potential $A_\mu$ and the tensor $F^{\mu\nu}$ defined by \cite{LL2V1975} % LL eq.(^26.5) D. - eq.(^ 23.7) i.e. neg. sign diff with LL
% We will follow LL, check agains D with the neg. showing up in some cases
%
%
%
\begin{eqnarray} 
F_{\mu\nu} =   \partial _\nu A_\mu - \partial_\mu A_\nu.
\label{Maxd2}
\end{eqnarray}
This lets us rewrite Maxwells equations in 4-D form as
\begin{eqnarray} 
F_{\mu\nu,\sigma} +F_{\nu\sigma,\mu} + F_{\sigma\mu,\nu}  = 0,
\label{Maxd3}
\end{eqnarray}
 for eqs. (\ref{Maxd1a}) and (\ref{Maxd1b}), while eqs. (\ref{Maxd1c}) and (\ref{Maxd1d}) become  %see LL eq. (30.2); neg of D eq. (23.11) 
\begin{eqnarray} 
F^{\mu\nu},_\nu  =  - 4 \pi J^\mu,
\label{Maxd4}
\end{eqnarray}
where $J^\mu$ is the 4-D current with $J^0 = \rho$.
Eq. (\ref{Maxd3}), looks like the Bianchi identities in GR \cite{Wei1972}.  These equations give  important differential identities that the field tensor $F^{\mu\nu}$ must obey.  As such, these equations give the covariant relation between the fields that must be satisfied in any frame of reference.    

The equations of motion for the electromagnetic interaction can be found from HP \cite{Dir1975}.  We write  
\begin{eqnarray}
S = S_m + S_f + S_{mf},
\label{Maxd5}
\end{eqnarray}
where $S_m$ is for the particles only,  $S_f$ is the action from the fields, and $S_{mf}$ is the interaction term.  For a single particle, we have 
\begin{eqnarray}
S_m = - m \int ds.
\label{Maxd5a}
\end{eqnarray}
If we have more than one particle, then we  would  sum over all particles.

For the fields we will have % LL pp 80; D eq.(^28.1) 
\begin{eqnarray}
\mathscr{L}_f =  - \frac{1}{16 \pi} F_{\mu\nu}F^{\mu\nu},
\label{Maxd6a}
\end{eqnarray}
and
\begin{eqnarray}
 S_f =  - \frac{1}{16 \pi} \int F_{\mu\nu}F^{\mu\nu} d^4x.
\label{Maxd6}
\end{eqnarray}
The interaction term for a single particle can be written as \cite{Dir1975} % D eq.(29.1)
\begin{eqnarray}
 S_{mf} = -  e \int A_\mu dx^\mu = - e \int A_\mu u^\mu ds,
\label{Maxd7}
\end{eqnarray}
where $e$ is the charge and $u^\mu$ is the particles velocity and the integration is along the the world line.  For a continuous distribution of charge carrying matter, this becomes
\begin{eqnarray}
 S_{mf} = - e \int A_\mu u^\mu ds \Rightarrow   - \int A_\mu J^\mu d^4x.
\label{Maxd8}
\end{eqnarray}
We wish to emphasise that the fields $A^\mu$ and $F^{\mu\nu}$ refer to both the external fields plus the field produced by the particles themselves.

We can get the equations of motion by applying HP to the total action of eq. (\ref{Maxd6}) plus eq. (\ref{Maxd8})
\begin{eqnarray}
\delta S = - \frac{1}{16\pi} \delta \int F_{\mu\nu}F^{\mu\nu} d^4x - \delta \int A_\mu J^\mu d^4x = 0.
\label{Maxd9}
\end{eqnarray}
We look at the change in the  action for a variation of the fields $A_\mu$.   In the first integral we use
\begin{eqnarray}
\delta F_{\mu\nu}F^{\mu\nu} = 4 F^{\mu\nu}  \delta A_{\mu,\nu},
\label{Maxd10}
\end{eqnarray}
and find
\begin{eqnarray}
\delta S =  \int \left\{\frac {1}{4 \pi} F^{\mu\nu} \delta A_{\mu,\nu}, + J^\mu \delta A_\mu \right\} d^4x = 0.
\label{Maxd11}
\end{eqnarray}
Integrating by parts and using the BC, we find
\begin{eqnarray}
\delta S =  \int \left\{ \frac{1}{4 \pi} F^{\mu\nu},_\nu + J^\mu \right\} \delta A_\mu  d^4x = 0.
\label{Maxd12}
\end{eqnarray}
Setting the bracket equal to zero gives us eq. (\ref{Maxd4}).  
%%%%%%%%%%%%%%%%%%%%%%%%%%%%%%%%%%%%%
%%%%%%%%%%%%%%%%%%%%%%%%%%%%%%%%%%%%%%
%
%
%
%
%
\subsection{Stress-Energy Tensor for Electric Fields}
%
%
%%%%%%%%%%%%%%%%%%%
%

We next find the  stress-energy tensor for the free field case.  We start with our equation for the Noether current, eq. (\ref{nNT3})
\begin{eqnarray}
j^\mu_i = 
    - \sum_s 
  \Bigg\{ \bigg{[}  
   \mathscr{L} \delta^\mu_\gamma
 -  \frac{\partial \mathscr{L}}{\partial  \phi_{s,\mu}} { \phi_{s, \gamma}} 
 \bigg{]}
  \frac{\partial x^\gamma}{\partial \omega_i}  \nonumber \\ + \frac{\partial \mathscr{L}}{\partial \phi_{ s,\mu}} \frac{\partial \phi_s}{\partial \omega_i}
  - \frac{\partial  \Gamma^\mu}{\partial  \omega_i} \Bigg\}.
\label{Maxd13}
\end{eqnarray}
with the variation of the coordinates given by eq. (\ref{dx11})
\begin{eqnarray}
\frac{\partial x^\gamma}{\partial \omega_i} = \delta_i^\gamma.
\label{Maxd14}
\end{eqnarray}
Using the Lagrangian given by eq. (\ref{Maxd6a}) and keeping the last term in eq. (\ref{Maxd13}) as \cite{LL2V1975n5} % LL, pp. 81
\begin{eqnarray}
\frac{\partial \Gamma^\mu}{\partial \omega_i} = - \frac{1}{4 \pi} \partial_\nu(A_\gamma F^{\mu\nu}) \delta^\gamma_i,
\label{Maxd15}
\end{eqnarray}
we find 
\begin{eqnarray}
j^\mu_i = \frac{1}{4 \pi} (
    \frac{1}{4} F_{\alpha\beta} F^{\alpha \beta} \delta^\gamma_i 
  - A_{\alpha,i} F^{\mu \alpha} +  A_{i, \nu} F^{\mu\nu}),
\label{Maxd16}
\end{eqnarray}
where we have used that for free fields
\begin{eqnarray}
  \partial_\nu F^{\mu\nu} = 0.
\label{Maxd17}
\end{eqnarray}
We may combine the last two terms to find % LL eq. (33.1) D. eq. (28.3) 
\begin{eqnarray}
T^{\mu\nu} =  \frac{1}{4 \pi} (
   - F^\nu_{~\alpha} F^{\mu \alpha}
    + \frac{1}{4} F_{\alpha\beta} F^{\alpha \beta} g^{\mu\nu}) .
\label{Maxd18}
\end{eqnarray}
For the more general case of including charged matter, the total stress-energy tensor will be conserved \cite{LL2V1975n5}.  

%%%%%%%%%%%%%%%%%%%%%%%%%%%%%%%%%%%%%%%%%%%
%%%%%%%%%%%%%%%%%%%%%%%%%%%%%%%%%%%%%%%%%%%
%
%%%%%%%%%%%%%%%%%%%%%%%%%%%%%%%%%%%%%%%%%%%%
%
%
\section{ Preliminary Considerations in General Relativity}
\label{GRPC}
%
%
%%%%%%%%%%%%%%%%%%%%%%%%%%%%%%%%%%%%%%%%%%%%%
%
%
We will look at the application of HP and NT in General Relativity. In this section, we set up equations by comparing to the Special Relativity case.

%%%%%%%%%%%%%%%%%%%%%%%%%%%%%%%%%%%%%
%%%%%%%%%%%%%%%%%%%%%%%%%%%%%%%%%%%%%%
%
%
%
%
%
\subsection{Equation of Motion }
%
%
%%%%%%%%%%%%%%%%%%%
%
 First we will derive the geodesic equation.  We find the equations of motion by applying HP to a line element.  For a point particle of mass $m$, we start with the variation of the action
\begin{eqnarray}
\delta S =  m~ \delta \int  {ds},
\label{GRa1}
\end{eqnarray}
where $ds$ is the line element, given by 
\begin{eqnarray}
ds^2 = g_{\mu \nu} dx^\mu dx^\nu.
\label{GRa2}
\end{eqnarray}
So we find
\begin{eqnarray}
\delta ds^2 = 2 ds \delta ds = \delta (g_{\mu \nu} dx^\mu dx^\nu)~~~~~~~~~~~~ \nonumber \\
= dx^\mu dx^\nu \frac{\partial g_{\mu \nu}}{\partial x^\gamma} \delta x^\gamma + 2  g_{\mu \nu} dx^\mu d \delta x^\nu.
\label{GRa3}
\end{eqnarray}
Subbing in above gives
\begin{eqnarray}
\delta S =  m 
\int \Bigg\{ \frac{1}{2} \frac{dx^\mu}{ds} \frac{dx^\nu}{ds} 
 \frac{\partial g_{\mu \nu}}{\partial x^\gamma} \delta x^\gamma \nonumber \\ + 
 g_{\mu \nu} \frac {dx^\mu}{ds} \frac {d \delta x^\nu}{ds} \Bigg\} ds.
\label{GRa4}
\end{eqnarray}
Integrating the second term by parts, gives
\begin{eqnarray}
\delta S =  m 
\int \Bigg\{ \frac{1}{2} \frac{dx^\mu}{ds} \frac{dx^\nu}{ds} 
 \frac{\partial g_{\mu \nu}}{\partial x^\gamma}  \nonumber \\ - 
 \frac{d}{ds} \left( g_{\mu \gamma} \frac {dx^\mu}{ds} \right) \Bigg\}  \delta x^\gamma ds
\label{GRa5}
\end{eqnarray}
Where we have changed the indexing in the last term so we can set the bracket equal to zero.
We replace the $dx^\mu/ds = u^\mu$ the velocity and expand out the derivative to find
\begin{eqnarray}
\frac{1}{2} u^\mu u^\nu 
 \frac{\partial g_{\mu \nu}}{\partial x^\gamma}  - 
  g_{\mu \gamma} \frac {d u^\mu}{ds}  - \frac{1}{2} \frac{ \partial  g_{\mu \gamma}}{\partial x^\nu} u^\nu u^\mu \nonumber \\
-  \frac{1}{2} \frac{ \partial  g_{\nu \gamma}}{\partial x^\mu} u^\nu u^\mu = 0.
\label{GRa6}
\end{eqnarray}
Using the definition of the Christoffel symbol \cite{LL2V1975n3}
\begin{eqnarray}
\Gamma^\alpha_{\mu \nu} = \frac{1}{2} g^{\alpha \gamma} \left(
   + \frac{ \partial  g_{ \gamma \mu}}{\partial x^\nu} 
+  \frac{ \partial  g_{ \gamma \nu}}{\partial x^\mu}
 - \frac{\partial g_{\mu \nu}}{\partial x^\gamma} 
  \right),
\label{GRa7}
\end{eqnarray}
we find the geodesic equation
\begin{eqnarray} 
 \frac {d u^\alpha}{ds}
   + \Gamma^\alpha_{\mu\nu} u^\nu u^\mu 
 = 0.
\label{GRa8}
\end{eqnarray}
This can be rewritten as 
\begin{eqnarray} 
 \frac {d^2 x^\alpha}{ds^2}
   = - \Gamma^\alpha_{\mu\nu} \frac{d x^\nu}{ds} \frac{d x^\mu}{ds}.
\label{GRa9}
\end{eqnarray}
Comparing to Newtonian mechanics,  the LHS gives the acceleration and the RHS plays the role of the force due to gravity.  With no gravity, i.e. space is flat, $\Gamma$ is zero and the equation can be integrated to give a straight line.  

The situation is different for light.  For light, we replace $u^\alpha$ in eq. (\ref{GRa6}) with $k^\alpha$, the propagation vector, and find \cite{LL2V1975n4} 
\begin{eqnarray} 
 \frac {d k^\alpha}{d\lambda}
   + \Gamma^\alpha_{\mu\nu} k^\nu k^\mu 
 = 0,
\label{GRa10}
\end{eqnarray}
where $\lambda$ is some parameter along the curve.  Since light will still propagate along a null vector, $k^\mu$ must still satisfy eq. (\ref{SRa2}).  With the definition, eq (\ref{SRa3}), we now find the eikonal equation \cite{LL2V1975n4} 
\begin{eqnarray} 
g^{\mu\nu} \partial _\mu \Psi \partial_\nu \Psi = 0.
\label{GRa11}
\end{eqnarray}
Thus we see, comparing to eq. (\ref{SRa4}), the flat space Minkowskian metric $\eta^{\mu\nu}$ is replaced by the metric for curved space $g^{\mu \nu}$.
%%%%%%%%%%%%%%%%%%%%%%%%%%%%%%%%%%%%%
%%%%%%%%%%%%%%%%%%%%%%%%%%%%%%%%%%%%%%
%
%
%
%
%
\subsection{Electric Fields in Curved Space-Time}
%
%
%%%%%%%%%%%%%%%%%%%
%
In this section the reader is referred to Dirac's book, Ref.\cite{Dir1975}.  The first set of field equations can be taken over to general relativity by using the "," goes to ";", i.e. replacing the partial derivatives by a covariant one.  Thus, eq. (\ref{Maxd3}) becomes
\begin{eqnarray} 
F_{\mu\nu;\sigma} +F_{\nu\sigma;\mu} + F_{\sigma\mu;\nu} = 0  \nonumber \\ = 
F_{\mu\nu,\sigma} +F_{\nu\sigma,\mu} + F_{\sigma\mu,\nu}.
\label{GRMaxd1}
\end{eqnarray}
 This is because the covariant derivatives adds two  correction terms to the partial derivatives and when all three $F$'s are added together, these correction tems cancel.  
So we see that these equations are the same.  

For our second set of equations, we have from eq. (\ref{Maxd4}), % LL eq. (90.6)  neg of D. eq.(23.13) 
\begin{eqnarray} 
F^{\mu\nu}_{~~;\nu}  =  - 4 \pi J^\mu.
\label{GRMaxd2}
\end{eqnarray}
Using the identity for any antisymmetric tensor $F^{\mu\nu}$ 
\begin{eqnarray} 
F^{\mu\nu}_{~~;\nu} \sqrt{-g} =  ( F^{\mu\nu} \sqrt{-g})_{,\nu}.
\label{GRMaxd3}
\end{eqnarray}
we can write this as % neg D. eq. (^ 24.1) 
\begin{eqnarray} 
 ( F^{\mu\nu} \sqrt{-g})_{,\nu}  =  - 4 \pi J^\mu \sqrt{-g}.
\label{GRMaxd4}
\end{eqnarray}
From this equation, one can see that % neg D. eq. (^ 24.1) 
\begin{eqnarray} 
 ( F^{\mu\nu} \sqrt{-g})_{,\mu \nu}  =  - 4 \pi (J^\mu \sqrt{-g})_{,\mu} = 0.
\label{GRMaxd5}
\end{eqnarray}
This is our continuity  equation that leads to the conservation of charge.  
%%%%%%%%%%%%%%%%%%%%%%%%%%%%%%%%%%%%%
%%%%%%%%%%%%%%%%%%%%%%%%%%%%%%%%%%%%%%
%
%
%
%
%
\subsection{Gravity and an Ideal Fluid or Gas}
%
%
%%%%%%%%%%%%%%%%%%%
%
In GR we write the action with four terms,

\begin{eqnarray}
S = S_g + S_m + S_f + S_{mf}
\label{GRMa1}
\end{eqnarray}
This is the same as in SR, eq. (\ref{Maxd5}), except we now have $S_g$ for the action of the metric that gives rise to gravity.  
 Einstein's equations for gravity can be found by starting with the Lagrangian density
\begin{eqnarray}
\mathscr{L}_g = (2 \kappa)^{-1} R \sqrt{-g},
\label{GRMa1a}
\end{eqnarray}
 and  varying the metric $g^{\mu\nu}$ to find\cite{LL2V1975} 
\begin{eqnarray}
\delta S_g = (2 \kappa)^{-1}
\delta \int   R   \sqrt {-g}  d^4x \nonumber \\ = (2 \kappa)^{-1}
{\int \left(  R^{ \mu\nu} - \frac{1}{2}  {g}^{\mu \nu}  R \right)  \sqrt {-g}~  \delta g_{\mu\nu}  d^4x} = 0,
\label{GRMa2}
\end{eqnarray}
where in our units $\kappa = 8 \pi G$, $G$ being Newton's gravitational constant.  This gives us for the gravitational field
\begin{eqnarray}
\psi^{\mu\nu}_g = (2\kappa)^{-1}
 \sqrt{-g} ( R^{ \mu\nu} - \frac{1}{2}  {g}^{\mu \nu}  R ) \nonumber \\ \equiv  (2\kappa)^{-1} \sqrt{-g} G^{\mu\nu}.
\label{GRMa2A}
\end{eqnarray}

We will always use the Lagrangian density $\mathscr{L}$ as a scalar.  The definition of a scalar can change depending on the type of space in which we are working in.  The main difference between SR and GR is the inclusion of $\sqrt{-g}$ in GR, which is identically equal to one in SR.
Setting $\psi$ to zero, the parenthesis is the Einstein equation for free space.  The constant $\kappa$ is to set the field strength for the gravitational interaction.  Usually one sets $\kappa = 8 \pi G$ where $G$ is the gravitational constant.  

We can add to this an area of space where there is an ideal gas or fluid.   We write the action as  
\begin{eqnarray}
 S_m = 
  \int  \mathscr{L}_m  d^4x =  \int   p  \sqrt {-g}  d^4x, 
\label{GRMa3}
\end{eqnarray}
 the matter Lagrangian, for an ideal fluid, is given by $p \sqrt{-g}$, $p$ being the pressure of the fluid.  When taking the variation of $p$, we base our procedure on the work of Bernard F. Schutz, Jr. \cite{Sch1970}.  We begin by writing that
\begin{eqnarray}
 p = p(\mu,  \mathbb{S}), 
\label{GRMa4}
\end{eqnarray}
where $\mu$ is the specific inertial mass and $\mathbb{S}$ is the specific entropy.  We also have the following relations, 
\begin{eqnarray}
 \rho = \rho_0 (1 +  \Pi), 
\label{GRMa5}
\end{eqnarray}
where $\rho$ density of total mass-energy, $\rho_0$ is the density of rest mass, and $\Pi$ is the specific internal energy, and the specific inertial mass is defined by 
\begin{eqnarray}
 \mu = (\rho + p)/\rho_0. 
\label{GRMa6}
\end{eqnarray}
Looking at the variation of the fluid with respect to the metric, we have
\begin{eqnarray}
 \delta S_m = 
\delta \int   p  \sqrt {-g}  d^4x~~~~~~~~~~~~~~~~~~~~~~~~~  \nonumber \\ =
 \int   \sqrt {-g}~ \delta  p ~  d^4x +  \int   p ~ \delta  \sqrt {-g} ~ d^4x.
\label{GRMa7}
\end{eqnarray}
Using the first and second laws of thermodynamics, we can get an expression for the variation of the pressure, staring with 
\begin{eqnarray}
 dp =\rho_0 d\mu - \rho_0 T d\mathbb{S}, 
\label{GRMa8}
\end{eqnarray}
 where $T$ is the temperature, we will have
\begin{eqnarray}
 \delta p =\rho_0 \delta\mu - \rho_0 T \delta\mathbb{S} =
 \rho_0 \frac{\partial \mu}{\partial g^{\mu\nu}} \delta g^{\mu\nu}, 
\label{GRMa9}
\end{eqnarray}
since varying the metric will be adiabatic, giving $\delta \mathbb{S} = 0$.  In order to get an expression for $\partial\mu/\partial g^{\mu\nu}$, we write $\mu$ is terms of the velocity vector $U^\mu$.  This vector is defined as
\begin{eqnarray}
 U^\mu =
  \frac{dx^\mu}{ds}, 
\label{GRMa10}
\end{eqnarray}
which has the normalisation
\begin{eqnarray}
 U^\mu U_\nu = 1. 
\label{GRMa11}
\end{eqnarray}
This vector has many uses, for example, the conservation of mass can be written as 
\begin{eqnarray}
\partial_\mu (\rho_0 U^\mu \sqrt{-g}) = 0, 
\label{GRMa12}
\end{eqnarray}
or 
\begin{eqnarray}
 ( \rho_0 U^\mu)_{;\mu}  = 0.
\label{GRMa13}
\end{eqnarray}
Consider writing an expression for $U^\mu$ as
\begin{eqnarray}
  U^\mu  = \mu^{-1} G^\mu.
\label{GRMa14}
\end{eqnarray}
where $G^\mu$ is not a function of the metric.  In general $G^\mu$ can be a function of other variables of the problem, see Ref. \cite{Sch1970} for examples.  Subbing this expression into the equation of normalisation, eq. (\ref{GRMa11}) we have 
\begin{eqnarray}
\mu^2 =  g^{\mu\nu} G_\mu G_\nu.
\label{GRMa15}
\end{eqnarray}
Differentiating eq. (\ref{GRMa15}) with respect to $g^{\mu\nu}$ 
\begin{eqnarray}
2 \mu \frac{\partial \mu}{\partial g^{\mu \nu}} =  G_\mu G_\nu.
\label{GRMa16}
\end{eqnarray}
or 
\begin{eqnarray}
 \frac{\partial \mu}{\partial g^{\mu \nu}} = \frac{\mu}{2}  U_\mu U_\nu.
\label{GRMa17}
\end{eqnarray}
We can sub this in to eq. (\ref{GRMa7}) to find
\begin{eqnarray}
 \delta S_m = 
  \int   \sqrt {-g}~ \rho_0 \frac{\mu}{2}  U_\mu U_\nu \delta g^{\mu\nu}   d^4x \nonumber \\ 
 + \int   p ~ \delta  \sqrt {-g} ~ d^4x.
\label{GRMa18}
\end{eqnarray}
Using the identities \cite{Dir1975}
\begin{eqnarray}
 \delta \sqrt{-g} 
  = -\frac{1}{2\sqrt{-g}} \delta g
  =   \frac{1}{2}\sqrt{-g}  g^{\alpha\beta} \delta g_{\alpha\beta}
\label{GRMaxe4}
\end{eqnarray}
and 
\begin{eqnarray}
 \delta g^{\mu\nu}
  =  -  g^{\mu\alpha}  g^{\nu\beta} \delta g_{\alpha\beta}.
\label{GRMaxe5}
\end{eqnarray}
we have
\begin{eqnarray}
 \delta S_m = 
  - \frac{1}{2} \int  \sqrt {-g}~ \rho_0 \mu  U_\mu U_\nu  g^{\mu\alpha}  g^{\nu\beta} \delta g_{\alpha\beta}   d^4x  \nonumber \\
 +  \int   p ~ \frac{1}{2}\sqrt{-g}  g^{\alpha\beta} \delta g_{\alpha\beta} ~ d^4x.
\label{GRMa19}
\end{eqnarray}
Subbing in for $\mu$, this reduces to 
\begin{eqnarray}
 \delta S_m = 
  - \frac{1}{2} \int \left[   (\rho + p)  U^\alpha U^\beta   
 -   p ~   g^{\alpha\beta} \right] \nonumber \\ \times \sqrt {-g} \delta g_{\alpha\beta} ~ d^4x,
\label{GRMa19b}
\end{eqnarray}
from which we identify the usual expression for the stress-energy tensor for an ideal fluid 
\begin{eqnarray}
 T^{\mu\nu}_m = 
    \left[   (\rho + p)  U^\mu U^\nu   
 -   p ~   g^{\mu\nu} \right].
\label{GRMa20}
\end{eqnarray}
 Einstein's field equations for this case becomes
\begin{eqnarray}
  R^{ \mu\nu} - \frac{1}{2}  {g}^{\mu \nu}  R = \kappa T^{\mu\nu}_m  .
\label{GRMa21}
\end{eqnarray}
%
%
%%%%%%%%%%%%%%%%%%%%%%%%%%%%%%%%%%%%%
%%%%%%%%%%%%%%%%%%%%%%%%%%%%%%%%%%%%%%
%
%
%
%
%
\subsection{Stress-Energy Tensor for Electric Fields}
%
%
%%%%%%%%%%%%%%%%%%%
%
We can add to eq. (\ref{GRMa21}) the contribution of the electric fields in free space.  This can be done by adding the electromagnetic stress-energy tensor to the RHS or eq. (\ref{GRMa21}).

We look at taking the variation of our electromagnetic field tensor to find the stress-energy tensor.   Similarly to Special Relativity, we  write the Lagrangian for the electric fields as \cite{Dir1975n1}
\begin{eqnarray}
\mathscr{L}_f =  - \frac{1}{16 \pi} F_{\mu\nu}F^{\mu\nu} \sqrt{-g},
\label{GRMaxe1}
\end{eqnarray}
where the factor of $\sqrt{-g}$ needs to be included in order for the Lagrangian to be a scalar.  Thus, the action becomes
\begin{eqnarray}
 S_f =  - \frac{1}{16 \pi} \int F_{\mu\nu}F^{\mu\nu} \sqrt{-g} d^4x \nonumber \\
  = - \frac{1}{16 \pi} \int F_{\mu\nu}F_{\alpha\beta}g^{\mu\alpha} g^{\nu\beta} \sqrt{-g} d^4x.
\label{GRMaxe2}
\end{eqnarray}
We next vary the $g_{\mu\nu}$ only.  So we will have
\begin{eqnarray}
 \delta( F_{\mu\nu}F^{\mu\nu} \sqrt{-g} )
  \nonumber =  \delta(F_{\mu\nu}F_{\alpha\beta}g^{\mu\alpha} g^{\nu\beta}\sqrt{-g} ) \\
  =   F_{\mu\nu}F^{\mu\nu}\delta \sqrt{-g} + F_{\mu\nu}F_{\alpha\beta} \sqrt{-g} \delta(g^{\mu\alpha} g^{\nu\beta})
\label{GRMaxe3}
\end{eqnarray}
Using the identities eq. (\ref{GRMaxe4}) and eq. (\ref{GRMaxe5})
we find 
\begin{eqnarray}
 \delta( F_{\mu\nu}F^{\mu\nu} \sqrt{-g} )
  =  \frac{1}{2} F_{\mu\nu}F^{\mu\nu}g^{\gamma\delta} \sqrt{-g} \delta g_{\gamma \delta} \nonumber \\
  - 2  F_{\mu\nu}F_{\alpha\beta} \sqrt{-g} g^{\mu\gamma} g^{\alpha \delta} g^{\nu\beta} \delta g_{\gamma \delta}.
\label{GRMaxe6}
\end{eqnarray}
Thus
\begin{eqnarray}
 &&\delta( F_{\mu\nu}F^{\mu\nu} \sqrt{-g} ) \nonumber \\
  &&= 2 ( \frac{1}{4}  F_{\mu\nu}F^{\mu\nu}g^{\gamma\delta} 
  -  F^{\gamma}_{~\beta}F^{\delta\beta})\sqrt{-g} \delta g_{\gamma \delta}.
\label{GRMaxe7}
\end{eqnarray}
The parenthesis is the stress-energy tensor for the electromagnetic fields times four pi.  So we find for the $\delta S_f$ 
\begin{eqnarray}
 \delta S_f =  - \frac{1}{2} \int  T^{\mu\nu} \sqrt{-g} ~\delta g_{\mu\nu} d^4x,
\label{GRMaxe8}
\end{eqnarray}
from which we deduce 
\begin{eqnarray}
 \psi^{(f)}_{\mu\nu} = \frac{\delta \mathscr{L}_f}{\delta g^{\mu\nu} }
 = -\frac{1}{2} T^{(f)}_{\mu\nu} \sqrt{-g},
\label{GRMaxe9}
\end{eqnarray}
where the subscript $(\mu\nu)$ on $\psi$ is an index for each $g^{\mu\nu}$.

We find that combining this with the gravitational field, $\delta S_g + \delta S_f$, gives
\begin{eqnarray}
  R^{ \mu\nu} - \frac{1}{2}  {g}^{\mu \nu}  R = \kappa T^{\mu\nu}_f  .
\label{GRMaxe10}
\end{eqnarray}
This is the Einstein equation for a matter free region of space with electromagnetic fields.  

Note that if we were to set $\delta S_f = 0$ this would give 
\begin{eqnarray}
 T^{\mu\nu}_f  = 0~~ {\rm{or}}~~  F^{\mu\nu}  = 0.
\label{GRMaxe11}
\end{eqnarray}
Thus every element of $F^{\mu\nu}$ would be zero.  This has been described as  disastrous \cite{BroB2002}, however it should be considered a reality check.  It says that a region of space that has no gravitation field, must have a zero electromagnetic field.  This makes sense because all electromagnetic fields are generated by charged particles and the particles must also generate gravitational fields.  To say that there is no gravitational fields, is to say that there are no particles,  thus no source terms for an electromagnetic field, giving $F^{\mu\nu} = 0$.
%%%%%%%%%%%%%%%%%%%%%%%%%%%%%%%%%%%%%
%%%%%%%%%%%%%%%%%%%%%%%%%%%%%%%%%%%%%%
%
%
%
%
%
\subsection{Maxwell's Equations }
%
%
%%%%%%%%%%%%%%%%%%%
%
We now derive Maxwell's Equations from H.P.  We will need expressions for the fields $S_f$ and for the interaction term $S_{mf}$.  To eq. (\ref{GRMaxe2})  we add the interaction term from eq. (\ref{Maxd7}) giving 
\begin{eqnarray}
 S_f + S_{mf} =  - \frac{1}{16 \pi} \int F_{\mu\nu}F^{\mu\nu} \sqrt{-g} d^4x \nonumber \\
  - \int \sigma A_\mu U^\mu \sqrt{-g} d^4x.
\label{GRMax1}
\end{eqnarray}
We vary this action with respect to $A_\mu$.  For the field term, we find (see \cite{LL2V1975}, pp 74)
\begin{eqnarray}
\delta F_{\mu\nu}F^{\mu\nu} 
  = - 4  F^{\mu\nu} \delta A_{\mu,\nu}.
\label{GRMax2}
\end{eqnarray}
Integrating by parts gives
\begin{eqnarray}
 \delta S_f  =  \frac{1}{4 \pi} \int F^{\mu\nu} \sqrt{-g} \delta A_{\mu,\nu} d^4x \nonumber \\
  = \frac{1}{4 \pi}  \int \big[(F^{\mu\nu} \sqrt{-g} \delta A_{\mu})_{,\nu}   \nonumber \\  - (F^{\mu\nu} \sqrt{-g})_{,\nu} \delta A_{\mu} \big] d^4x,
\label{GRMax3}
\end{eqnarray}
where the first term integrates to zero because of the b.c.  We rewrite the derivative in the second term as a covariant derivative using eq. (\ref{GRMaxd3}), thus,
\begin{eqnarray}
 \delta S_f  
  = - \frac{1}{4 \pi} \int  F^{\mu\nu}_{~;\nu} \sqrt{-g}  \delta A_{\mu}  d^4x.
\label{GRMax4}
\end{eqnarray}
For the interaction term, we have 
\begin{eqnarray}
\delta S_{mf} =  
  - \int \sigma U^\mu \sqrt{-g} \delta A_\mu d^4x \nonumber \\ =  - \int J^\mu \sqrt{-g}  \delta A_\mu d^4x,
\label{GRMax5}
\end{eqnarray}
where we have identified the current as
\begin{eqnarray}
  J^\mu =  \sigma U^\mu.
\label{GRMax6}
\end{eqnarray}
Combining the two integrals we have
\begin{eqnarray}
 \delta S_f  + \delta S_{mf} 
  = -  \int \left(\frac{1}{4 \pi} F^{\mu\nu}_{~;\nu} + J^\mu\right) \sqrt{-g}  \delta A_{\mu}  d^4x.
\label{GRMax7}
\end{eqnarray}
and setting the integral equal to zero, we find
\begin{eqnarray}
  F^{\mu\nu}_{~;\nu} = -  4 \pi J^\mu.
\label{GRMax8}
\end{eqnarray}
This is eq. (\ref{GRMaxd2}) that was found, from the SR equation, by using the {\textquotedblleft ~, \textquotedblright} goes to {\textquotedblleft ~; \textquotedblright} ~procedure.
%
%
%%%%%%%%%%%%%%%%%%%%%%%%%%%%%%%%%%%%%%%%%%%
%%%%%%%%%%%%%%%%%%%%%%%%%%%%%%%%%%%%%%%%%%%
%
%%%%%%%%%%%%%%%%%%%%%%%%%%%%%%%%%%%%%%%%%%%%
%
%
\section{ Noether's First Theorem in  General Relativity}
\label{GRNT1}
%
%
%%%%%%%%%%%%%%%%%%%%%%%%%%%%%%%%%%%%%%%%%%%%%
%
%
The problem of understanding the  conservation of energy in GR and in general for covariant theories derived from a variational principle, is what is behind Noether's investigation of such theories \cite{Brad2005}.   Very briefly,  in 1916, Einstein introduced a pseudotensor to accomplish energy conservation \cite{Ein1916}.  Hilbert's conjecture was also published in 1916 \cite{Hilb1916}.  Klein objected to the energy conservation treatment by Hilbert and Einstein, claiming that their conservation laws were meer mathematical identities, and wrote letters to both.  In response, a Hilbert letter to Klein in 1917, Hilbert said that he had asked for Noether's help  with the clarification of questions in the analytical treatment of his energy theorem, more than a year ago \cite{Klei1917}.     Noether's work resulted in her  1918 paper \cite{Noe1918} with the general theorems.  From 1917, Klein continues to work on the problem with the assistance of Noether and in 1918, Klein presented his paper that contained The Third Theorem \cite{Klei1918}, which is a rewrite of Noether's Second Theorem, applying it to energy conservation in GR.  For more detail, see Brading \cite{Brad2005} and works cited.        

%%%%%%%%%%%%%%%%%%%%%%%%%%%%%%%%%%%%%
%%%%%%%%%%%%%%%%%%%%%%%%%%%%%%%%%%%%%%
%
%
%
%
%
\subsection{Noether's First Theorem }
%
%
%%%%%%%%%%%%%%%%%%%
%
In this section, we apply Noether's First Theorem in GR \cite{BroB2002}.  We will look at applying it to the Lagrangian for the electromagnetic field, eq. (\ref{GRMaxe1}). We consider the usual infinitesimal  linear coordinate transformation, leading to eq. (\ref{dx11}),
\begin{eqnarray}
\frac{\partial  x^\gamma}{\partial  \omega_i} = \delta_i^\gamma.
\label{GRNT1a1}
\end{eqnarray}
By subbing into Noether's First Theorem, eq. (\ref{nNT1}) and using eq. (\ref{GRMaxe9}), we find
\begin{eqnarray}
 -\frac{1}{2} T_{\mu\nu} g^{\mu\nu}_{~,\alpha} \sqrt{-g} = \partial _\mu  j^\mu_\alpha.
\label{GRNT1a2}
\end{eqnarray}
To get an expression for the Noether currents, we sub into eq. (\ref{nNT3}), 
\begin{eqnarray}
j^\mu_\alpha =  - 
   \left(  
   \mathscr{L} \delta^\mu_\gamma
 -  \frac{\partial \mathscr{L}}{\partial  A_{\alpha,\mu}} { A_{\alpha, \gamma}} 
 \right),
\label{GRNT1a3}
\end{eqnarray}
and find
\begin{eqnarray}
j^\mu_\alpha = \frac{ \sqrt{-g} }{4 \pi}
   \left( 
   F^{\mu\beta} A_{\beta , \alpha}
 -  \frac{1}{4} \delta^\mu_\alpha F_{\rho\lambda} F^{\rho\lambda} 
 \right).
\label{GRNT1a4}
\end{eqnarray}
This is similar to what was found in eq. (\ref{Maxd16}).  We can use a similar procedure here to find
\begin{eqnarray}
j^\mu_\alpha =  \sqrt{-g} ~T^\mu_\alpha
 -  \frac{1}{4\pi} \partial_\beta ( \sqrt{-g} F^{\mu\beta} A_\alpha).
\label{GRNT1a5}
\end{eqnarray}
We see that the Noether current differs form $\sqrt{-g} T^\mu_\alpha$ by a divergence term, which will give zero contribution when subbed into eq. (\ref{GRNT1a2}). Dropping the second term, we have 
\begin{eqnarray}
\partial _\mu  ( \sqrt{-g} ~T^\mu_\alpha)
 -\frac{1}{2} T_{\mu\nu} g^{\mu\nu}_{~,\alpha} \sqrt{-g} = 0.
\label{GRNT1a6}
\end{eqnarray}
We recognise this as the covariant derivative \cite{Tol1950}, thus
\begin{eqnarray}
 T^\mu_{\nu ; \mu} = 0.
\label{GRNT1a7}
\end{eqnarray}
Thus, Noether's  First Theorem dose not lead to the conservation of anything, since we find a covariant derivative instead of a partial derivative is equal to zero.  Although we found this result for electric fields, the results are general and apply to all interactions.  Remember, that we have not considered internal degrees of freedom, such as spin.

%%%%%%%%%%%%%%%%%%%%%%%%%%%%%%%%%%%%%
%%%%%%%%%%%%%%%%%%%%%%%%%%%%%%%%%%%%%%
%
%
%
%
%
\subsection{Gravitational Energy }
%
%
%%%%%%%%%%%%%%%%%%%
%
To understand gravitational energy in GR, we must first understand the drastic change in the description of physics in going from SR to GR.  In SR, the canvas that we paint our equations onto, such as Minkowskian space-time, is flat, ridged and unchanging.  It matters not if we are describing a dust particle in Browning motion or a speeding asteroid heading straight to earth,  the space-time is unchanged as the events unfold.  In GR things change.  There are no ridged coordinate systems.  As events unfold, the coordinate system changes.  Even for our solar system, the radiation from the sun, changes the coordinates, as it's energy is lost to the system (although very little).   The only way to prevent change is to have a perfectly static case.

In SR, conservation of energy is stated with the equation \cite{Trau1962}
\begin{eqnarray}
  T^{\nu}_{\mu,\nu} = 0.
\label{GRE1}
\end{eqnarray}
In GR, this becomes \cite{LL2V1975,Dir1975} 
\begin{eqnarray}
\nonumber   T^{\nu}_{\mu;\nu}  =&& T^{\nu}_{\mu,\nu} -\Gamma^\alpha_{\mu\nu} T^\nu_\alpha
  + \Gamma^\nu_{\alpha\nu} T^\alpha_\mu \\
  =&& \frac{1}{\sqrt{-g}} \frac{\partial (T^\nu_\mu \sqrt{-g})}{\partial x^\nu} 
 - \frac{1}{2} \frac{\partial g_{\nu\alpha}}{\partial x^\mu} ~T^{\nu\alpha}
  = 0.
\label{GRE2}
\end{eqnarray}
It has been observed that \textquotedblleft this equation does not generally express any conservation law whatever\textquotedblright~\cite{LL2V1975}.  The problem is that the $T$'s are derived for some source of energy, such as electromagnetism and the derivative contains a term with $g$'s, that are associated with gravity.  Any energy conservation law therefore must include the contribution from gravity.  However, there is not a way to derive a stress energy tensor for the gravitational field since it has been replaced with the geometry of the metric.  We are left with an alternative formulation of the general form \cite{Trau1962} 
\begin{eqnarray}
  (T^\nu_\mu  + t^\nu_\mu )_{,\nu}= 0,
\label{GRE3}
\end{eqnarray}
where $ T^\nu_\mu$ is the total stress energy matrix from all source except gravity,  $ t^\nu_\mu$ is a pseudotensor that will account for the energy of gravity.  $ t^\nu_\mu$  has no tensorial properties.  If one was to solve eq. (\ref{GRE3}) for $ t^\nu_\mu$ and then solve
\begin{eqnarray}
  (T^{\nu\mu}  + t^{\nu\mu} )_{,\nu}= 0,
\label{GRE4}
\end{eqnarray}
for $ t^{\nu\mu}$ one should expect 
\begin{eqnarray}
 t^{\nu\mu} \neq  g^{\mu\alpha}  t^\nu_\alpha.
\label{GRE5}
\end{eqnarray}
In all that follows, $ T^\nu_\mu$ will not contain any contributions form gravity.  There are two approaches to finding the $t^{\nu\mu}$'s, either by fixing a coordinate system or by choosing a special vector field.

%%%%%%%%%%%%%%%%%%%%%%%%%%%%%%%%%%%%%
%%%%%%%%%%%%%%%%%%%%%%%%%%%%%%%%%%%%%%
%
%
%
%
%
\subsection{The Landau and Lifshitz Pseudotensor  }
%
%
%%%%%%%%%%%%%%%%%%%
%
We outline the presentation of Landau and Lifshitz \cite{LL2V1975n6} for the formulation of their pseudotensor for the gravitational energy.  We start by picking a coordinate system for which the first derivative of the metric is zero at some point.
\begin{eqnarray}
 \frac{\partial g_{\mu\nu}}{  \partial x^\alpha} = 0.
\label{GRE6}
\end{eqnarray}
So from eq. (\ref{GRE2}) we get 
\begin{eqnarray}
 \frac{\partial  T^{\nu}_{\mu}}{\partial x^\nu}  =0
  =  \frac{\partial  T^{\mu\nu}}{\partial x^\nu}.
\label{GRE7}
\end{eqnarray}
This allows us to write the following expression 
\begin{eqnarray}
 (-g)  T^{\mu\nu} = \frac{\partial h^{\mu\nu\alpha}}{\partial x^\alpha}.
\label{GRE8}
\end{eqnarray}
where 
\begin{eqnarray}
  h^{\mu\nu\alpha} = \frac{1}{4 \kappa} \frac{\partial}{\partial x^\beta}
  [(-g)(g^{\mu\nu}g^{\alpha\beta} - g^{\mu\alpha}g^{\nu\beta}).
\label{GRE9a}
\end{eqnarray}
We  find that $ h^{\mu\nu\alpha}$ is antisymmetric in $\nu$ and $\alpha$, so
\begin{eqnarray}
  h^{\mu\nu\alpha} = - h^{\mu\alpha\nu}.
\label{GRE9}
\end{eqnarray}
In this coordinate system, we can write
\begin{eqnarray}
 \frac{\partial h^{\mu\nu\alpha}}{\partial x^\alpha} -  (-g)  T^{\mu\nu} = 0.
\label{GRE10}
\end{eqnarray}
However, in an arbitrary coordinate system, this difference will not be zero.  We can write the difference as $(-g) t^{\mu\nu}$ leading to 
\begin{eqnarray}
 (-g)  ( T^{\mu\nu} + t^{\mu\nu} ) = \frac{\partial h^{\mu\nu\alpha}}{\partial x^\alpha},
\label{GRE11}
\end{eqnarray}
with the symmetry 
\begin{eqnarray}
 t^{\mu\nu}  = t^{\nu\mu}.
\label{GRE12}
\end{eqnarray}
We can use the Einstein Field Equation to get 
\begin{eqnarray}
 (-g) [ \frac{1}{2 \kappa} ( R^{\mu\nu} - \frac{1}{2} g^{\mu\nu} R) + t^{\mu\nu}  ] 
 = \frac{\partial h^{\mu\nu\alpha}}{\partial x^\alpha}.
\label{GRE13}
\end{eqnarray}
This will give us an expression for $t^{\mu\nu}$ in terms of the $g$'s \cite{LL2V1975},
\begin{eqnarray}
&&(-g) t^{\mu\nu} = \frac{1}{4 \kappa} \big\{
\textgoth{g}^{\mu\nu}_{~,\alpha} \textgoth{g}^{\alpha\beta}_{~,\beta} 
-\textgoth{g}^{\mu\alpha}_{~,\alpha} \textgoth{g}^{\nu\beta}_{~,\beta} 
 + \frac{1}{2} g^{\mu\nu} g_{\alpha\beta}  
 \textgoth{g}^{\alpha\gamma}_{~,\delta} \textgoth{g}^{\delta\beta}_{~,\gamma}  \nonumber \\
&& - g^{\mu\alpha}g_{\beta\gamma}
 \textgoth{g}^{\nu\gamma}_{~,\delta} \textgoth{g}^{\beta\delta}_{~,\alpha} 
- g^{\nu\alpha}g_{\beta\gamma}
 \textgoth{g}^{\mu\gamma}_{~,\delta} \textgoth{g}^{\beta\delta}_{~,\alpha} 
+ g^{\alpha\beta}g_{\gamma\delta}
 \textgoth{g}^{\mu\gamma}_{~,\alpha} \textgoth{g}^{\nu\delta}_{~,\beta} \nonumber \\
&&+ \frac{1}{8} (2 g^{\mu\alpha} g^{\nu\beta} - g^{\mu\nu} g^{\alpha\beta} ) \nonumber \\
&& \times (2 g_{\rho\eta} g_{\lambda\xi} - g_{\eta\lambda} g_{\rho\xi} )
 \textgoth{g}^{\rho\xi}_{~,\alpha} \textgoth{g}^{\eta\lambda}_{~,\beta} 
  \big\},
\label{GRE13b}
\end{eqnarray}
where $\textgoth{g}^{\mu\nu} = \sqrt{-g}{g}^{\mu\nu}$.  Now from eq. (\ref{GRE11}), we have
\begin{eqnarray}
 \frac{\partial}{\partial x^\nu} [ (-g)  ( T^{\mu\nu} + t^{\mu\nu} ) ] = 0.
\label{GRE14}
\end{eqnarray}
This is in the desired form.  From eq. (\ref{GRE14}) we can find  conservation laws in the usual way.  We have used the fixed coordinate method to find $ t^{\mu\nu}$. 

%%%%%%%%%%%%%%%%%%%%%%%%%%%%%%%%%%%%%
%%%%%%%%%%%%%%%%%%%%%%%%%%%%%%%%%%%%%%
%
%
%
%
%
\subsection{The Dirac Pseudotensor  }
%
%
%%%%%%%%%%%%%%%%%%%
%
 Dirac's work on his pseudotensor for the conservation of energy, is similar to the work of Schr{\"{o}}dinger \cite{Schr1950}. Although the notation is different and Dirac's initial formulation is more general,  the final equations are essentially the same.  This section follows Dirac \cite{Dir1975n2}. 

Dirac derives an alternate to eq. (\ref{GRE14}) by first defining
\begin{eqnarray}
 \sqrt{-g}  t_\mu^\nu   = \frac{\partial \mathscr{L}}{\partial g_{\alpha\beta,\nu}}  g_{\alpha\beta,\mu} 
 -g^\nu_\mu \mathscr{L},
\label{GRE15}
\end{eqnarray}
where $\mathscr{L}$ is the Lagrangian for gravity only.  By taking the partial of this equation with respect to $\nu$ and using the general expression for the derivative of $\mathscr{L}$ we find
\begin{eqnarray}
 (\sqrt{-g}  t_\mu^\nu)_{,\nu}   = \left[ \left( \frac{\partial \mathscr{L}}{\partial g_{\alpha\beta,\nu}}  \right)_{,\nu} 
 -\frac{ \partial \mathscr{L}}{\partial g_{\alpha\beta}} \right] g_{\alpha\beta,\mu}
\label{GRE16}
\end{eqnarray}
From eqs. (\ref{GRMaxe9}) and(\ref{GRMaxe10}) we recognise the square bracket as the Lagrange Expression for the $g_{\mu\nu}$'s and can write 
\begin{eqnarray}
 (\sqrt{-g}  t_\mu^\nu)_{,\nu}   = \frac{1}{2 \kappa}  \left( R^{\alpha\beta} 
 - \frac{1}{2} g^{\alpha\beta} R  \right) g_{\alpha\beta,\mu} \sqrt{-g},
\label{GRE17}
\end{eqnarray}
or
\begin{eqnarray}
 (\sqrt{-g}  t_\mu^\nu)_{,\nu}   = \frac{1}{2 }  T^{\alpha\beta} 
  g_{\alpha\beta,\mu} \sqrt{-g}.
\label{GRE18}
\end{eqnarray}
We use eq. (\ref{GRNT1a6}) 
\begin{eqnarray}
   (\sqrt{-g}  T_\mu^\nu)_{,\nu} = - \frac{1}{2 }  T^{\alpha\beta} 
  g_{\alpha\beta,\mu} \sqrt{-g},
\label{GRE19}
\end{eqnarray}
and subbing into eq. (\ref{GRE18}), gives us
\begin{eqnarray}
 [\sqrt{-g} ( t_\mu^\nu + T_\mu^\nu )]_{,\nu}   = 0.
 \label{GRE20}
\end{eqnarray}
This looks like a good conservation rule, however, $t_\mu^\nu$ is not a tensor.  This can be seen by noting that $t_\mu^\nu$  can be written as 
\begin{eqnarray}
 \sqrt{-g}  t_\mu^\nu   = \frac{\partial {L}}{\partial g_{\alpha\beta,\nu}}  g_{\alpha\beta,\mu} 
 -g^\nu_\mu {L},
\label{GRE21}
\end{eqnarray}
where $\sqrt{-g} L = \mathscr{L}$.  Since $L$ is not a scalar, $ t_\mu^\nu $ is not a tensor.

To find an explicit expression for $ t^\nu_\mu$, following Dirac \cite{Dir1975n2}, we rewrite eq. (\ref{GRE15}) as
\begin{eqnarray}
 \sqrt{-g}  t_\mu^\nu   = \frac{\partial {\mathscr{L}}}{\partial q_{n,\nu}}  q_{n,\mu} 
 -g^\nu_\mu {\mathscr{L}},
\label{GRE22}
\end{eqnarray}
where we have written $q_n,  n = 1, ... ,10$ for the ten independent $g_{\mu\nu}$'s.  We could write, in place of the $q$'s, any ten independent functions of the $q$'s, $Q_n$ and would still satisfy eq. (\ref{GRE21}),
\begin{eqnarray}
 \sqrt{-g}  t_\mu^\nu   = \frac{\partial {\mathscr{L}}}{\partial Q_{n,\nu}}  Q_{n,\mu} 
 -g^\nu_\mu {\mathscr{L}}.
\label{GRE23}
\end{eqnarray}
We continue by choosing
\begin{eqnarray}
 Q_n = g^{\mu\nu} \sqrt{-g} = \textgoth{g}^{\mu\nu}.
\label{GRE24}
\end{eqnarray}
Next, we will use an alternate  expression for $R$ to use in $\mathscr{L}$,
\begin{eqnarray}
 &&R = g^{\mu\nu} R_{\mu\nu} \nonumber \\ &&= g^{\mu\nu} ( \Gamma^\alpha_{\mu\alpha,\nu} -  \Gamma^\alpha_{\mu\nu,\alpha} 
 - \Gamma^\alpha_{\mu\nu} \Gamma^\beta_{\alpha\beta} + \Gamma^\alpha_{\mu\beta} \Gamma^\beta_{\nu\alpha} ),
\label{GRE25}
\end{eqnarray}
so we have
\begin{eqnarray}
 \mathscr{L} = \textgoth{g}^{\mu\nu} ( \Gamma^\alpha_{\mu\alpha,\nu} -  \Gamma^\alpha_{\mu\nu,\alpha} 
 - \Gamma^\alpha_{\mu\nu} \Gamma^\beta_{\alpha\beta} + \Gamma^\alpha_{\mu\beta} \Gamma^\beta_{\nu\alpha} ).
\label{GRE26}
\end{eqnarray}
The first two terms can be rewritten as total derivatives minus the difference (just like integration by parts). Thus,
looking just at the first two terms, we have
\begin{eqnarray}
 \textgoth{g}^{\mu\nu} ( \Gamma^\alpha_{\mu\alpha,\nu} -  \Gamma^\alpha_{\mu\nu,\alpha}) = 
  (\textgoth{g}^{\mu\nu}  \Gamma^\alpha_{\mu\alpha})_{,\nu} \nonumber \\
  - (\textgoth{g}^{\mu\nu} \Gamma^\alpha_{\mu\nu})_{,\alpha}
  - \textgoth{g}^{\mu\nu}_{~,\nu}  \Gamma^\alpha_{\mu\alpha} 
  + \textgoth{g}^{\mu\nu}_{~,\alpha}   \Gamma^\alpha_{\mu\nu}.
\label{GRE27}
\end{eqnarray}
The first two terms on the right of eq. (\ref{GRE27}) are in the form of a divergence  and will not contribute to the equations of motion.  They can be dropped.  So we write our effective Lagrangian as
\begin{eqnarray}
 \mathscr{L} = ( \Gamma^\nu_{\alpha\beta} - g^\nu_\beta \Gamma^\sigma_{\alpha\sigma})\textgoth{g}^{\alpha\beta}_{,\nu}  \nonumber \\
 -(\Gamma^\alpha_{\mu\nu} \Gamma^\beta_{\alpha\beta} 
 - \Gamma^\alpha_{\mu\beta} \Gamma^\beta_{\nu\alpha} )  \textgoth{g}^{\mu\nu}.
\label{GRE28}
\end{eqnarray}
Using this form of the Lagrangian, we find for eq. (\ref{GRE22})
\begin{eqnarray}
 2 \kappa \sqrt{-g}  t_\mu^\nu   =  ( \Gamma^\nu_{\alpha\beta} 
 - g^\nu_\beta \Gamma^\sigma_{\alpha\sigma})\textgoth{g}^{\alpha\beta}_{,\mu}
 -g^\nu_\mu {\mathscr{L}},
\label{GRE29}
\end{eqnarray}
where we have included the coupling with the $t^\nu_\mu$ term.
This gives us an expression for $t_\mu^\nu$ purely in terms of the metric.  This procedure  used the special vector field method.

%%%%%%%%%%%%%%%%%%%%%%%%%%%%%%%%%%%%%
%%%%%%%%%%%%%%%%%%%%%%%%%%%%%%%%%%%%%%
%
%
%
%
%
\subsection{Energy in Cosmology }
%
%
%%%%%%%%%%%%%%%%%%%
%

  An exception to the general result of no conservation rule for energy, are spaces that have constant curvature \cite{BroB2002}. The standard model in cosmology uses  the Robinson-Walker Metric 
\begin{eqnarray}
d\tau^2 =  dt^2
 - R^2(t) \Big\{ \frac{dr^2}{1 - kr^2} \nonumber \\ + r^2 d\theta^2 + r^2 sin^2 \theta ~d\phi^2 \Big\}.
\label{GRECs01}
\end{eqnarray}
In this metric, $R(t)$ is an unknown function of time and $k$ is a constant, that in the usual choice of unites, can have values of $+1$, $0$, or $-1$, which correspond to a closed, flat or open universe.  
The function $R(t)$ is called the cosmic scale factor and, for the closed universe, the radius of the universe.  

These are co-moving coordinates, and the average motion of the content of the universe with respect to these coordinates, is zero.  As such, we have that the velocity vector is $U^0 = 1$ and $U^i = 0$.  Using this in our geodesic equation, we have
\begin{eqnarray} 
 \frac {d u^\alpha}{ds}
   + \Gamma^\alpha_{00}  
 = 0.
\label{GRECs02a}
\end{eqnarray}
For our metric, $\Gamma^\alpha_{00} = 0$, so we find
\begin{eqnarray} 
 \frac {d u^\alpha}{ds}
 = 0,
\label{GRECs02}
\end{eqnarray}
that is that the velocity vector of the energy content $\rho$ only has local motion, but on the average does not move relative to the co-moving coordinates of the universe.

For this metric the 3-D subspace of the spacial part is maximally symmetric, and thus has constant curvature. This will allow us to find a conservation of energy equation.   We start with assuming that the content of the universe is a perfect fluid or gas.  The energy momentum tensor is given by eq. (\ref{GRMa20})
\begin{eqnarray}
 T^{\mu\nu}_m = 
    \left[   (\rho + p)  U^\mu U^\nu   
 -   p ~   g^{\mu\nu} \right],
\label{GRECs03}
\end{eqnarray}
Using this in the results of Noether's First Theorem, eq. (\ref{GRNT1a7}) we can write
\begin{eqnarray}
 0 = T^{\mu\nu}_ {; \mu} 
 = \frac{\partial T^{\mu\nu}}{\partial x^\mu} +\Gamma^\mu_{\mu\lambda}T^{\lambda\nu}
 +\Gamma^\nu_{\mu\lambda}T^{\mu\lambda} .
\label{GRECs04}
\end{eqnarray}
The $\Gamma$ terms are the contribution of gravity to the energy.  We can use the identity
\begin{eqnarray}
 \Gamma^\mu_{\mu\lambda}
= \frac{1}{\sqrt{-g}} \frac{\partial}{\partial x^\lambda} \sqrt{-g}
\label{GRECs05}
\end{eqnarray}
to rewrite this as
\begin{eqnarray}
 T^{\mu\nu}_ {; \mu} 
 =\frac{1}{\sqrt{-g}} \frac{\partial}{\partial x^\lambda} ( \sqrt{-g}T^{\lambda\nu})
 +\Gamma^\nu_{\mu\lambda}T^{\mu\lambda} .
\label{GRECs06}
\end{eqnarray}
 Subbing in for our energy momentum tensor from eq. (\ref{GRECs03}) an reducing we find
\begin{eqnarray}
 T^{\mu\nu}_ {; \nu} 
 =\frac{\partial p}{\partial x^\nu} g^{\mu\nu}
 + \frac{1}{\sqrt{-g}} \frac{\partial}{\partial x^\nu} 
 [ \sqrt{-g}     (\rho + p)  U^\mu U^\nu] \nonumber \\
 +\Gamma^\mu_{\nu\lambda}  (\rho + p)  U^\nu U^\lambda.
\label{GRECs07}
\end{eqnarray}
By subbing in for our metric $\Gamma^\alpha_{00} = 0$  and our velocity vector  $U^0 = 1$ and $U^i = 0$, we find that the spacial equations are trivial and the temporal equation reduces to 
\begin{eqnarray}
R^3(t) \frac{dp(t)}{dt}
= \frac{d}{dt} \{ R^3(t) [ \rho(t) +p(t) ] \}. 
\label{GRECs08}
\end{eqnarray}
 For the case that the pressure is very small compared to the density and can be ignored, we find
\begin{eqnarray}
 R^3(t)  \rho(t) = constant. 
\label{GRECs09}
\end{eqnarray}
This makes sense that the energy density would decrease as the volume of the universe increases, etc.  

%%%%%%%%%%%%%%%%%%%%%%%%%%%%%%%%%%%%%%%%%%%
%%%%%%%%%%%%%%%%%%%%%%%%%%%%%%%%%%%%%%%%%%%
%

%%%%%%%%%%%%%%%%%%%%%%%%%%%%%%%%%%%%%%%%%%%%
%
%
\section{ Noether's Second Theorem in  General Relativity}
\label{GRNT2}
%
%
%%%%%%%%%%%%%%%%%%%%%%%%%%%%%%%%%%%%%%%%%%%%%
%
%
In this section, we will look at the formulation of Noether's Second Theorem and The Third Theorem in General Relativity \cite{BroB2002}.

%%%%%%%%%%%%%%%%%%%%%%%%%%%%%%%%%%%%%
%%%%%%%%%%%%%%%%%%%%%%%%%%%%%%%%%%%%%%
%
%
%
%
%
\subsection{Noether's Second Theorem and the Bianchi Identity }
%
%
%%%%%%%%%%%%%%%%%%%
%
Before we write out Noether's Second Theorem, a bit of history.  When dealing with GR. it is found that all the Euler-Lagrange equations are not independent.  This appears to be first pointed out by Hilbert \cite{BroB2002}.  From 1915 to 1918 David Hilbert,  Hermann Weyl, Felix Klein and   Emmy Noether  investigated the variational approach to GR. This problem and related considerations were addressed rigorously and in general  in Noether's 1918 paper \cite{Noe1918}.  More recently, this issue has been noted, see for example Schr{\"{o}}dinger \cite{Schr1950}.  Dirac, seems to have rediscovered the problem \cite{Dir1975}, however, Noether's Second Theorem makes the issue, manifest.  

We start with Noether's Second Theorem, eq. (\ref{nNT20}), 
\begin{eqnarray}
\sum_s \psi_s a_{is} = 
\sum_s   \partial_\mu (\psi_s b^\mu_{is}).
\label{GRNT2a1}
\end{eqnarray}
We will rewrite it by separating out the metric contribution from the other fields \cite{BroB2002},
\begin{eqnarray}
 &&\psi_{\mu\nu} g^{\mu\nu}_{~;\alpha}  +  \sum_r \psi^r a_{is} \nonumber \\ &&= 2(g^{\mu\beta}\psi_{\mu\alpha} )_{;\beta} 
+ \sum_r   \partial_\mu (\psi^r b^\mu_{is}).
\label{GRNT2a2}
\end{eqnarray}
Several comments are necessary at this point.  The first term in eq. (\ref{GRNT2a2}) contains the covariant derivative of the metric.  We will investigate this by looking at \cite{MTW1973}
\begin{eqnarray}
 g_{\mu\nu;\alpha}   &&= 
g_{\mu\nu,\alpha} - \Gamma^\beta_{\mu\alpha} g_{\beta\nu}  - \Gamma^\beta_{\nu\alpha} g_{\beta\mu} \nonumber \\ &&=
g_{\mu\nu,\alpha} - \Gamma_{\nu\mu\alpha}  - \Gamma_{\mu\nu\alpha}.
\label{GRNT2a3}
\end{eqnarray}
The connection coefficients, in any basis, is written 
\begin{eqnarray}
 \Gamma_{\mu\nu\alpha}  = \frac{1}{2} (g_{\mu\nu,\alpha} +  g_{\mu\alpha,\nu} - g_{\nu\alpha,\mu} \nonumber \\
  + c_{\mu\nu\alpha} +  c_{\mu\alpha\nu} - c_{\nu\alpha\mu}).
\label{GRNT2a4}
\end{eqnarray}
where the $c$'s are called the commutation coefficients  for a basis $e_\alpha$ and are defined by \\
\begin{eqnarray}
  [e_\mu , e_\nu] = [\partial_\mu , \partial_\nu] 
   =  c_{\mu\nu}^{~~\alpha} e_\alpha,
\label{GRNT2a5}
\end{eqnarray}
with
\begin{eqnarray}
   c_{\mu\nu\alpha} =   g_{\alpha\beta} c_{\mu\nu}^{~~\beta}.
\label{GRNT2a6}
\end{eqnarray}
Equation (\ref{GRNT2a3}) is called the compatibility equation and if we have 
\begin{eqnarray}
 g_{\mu\nu;\alpha}   =  0,
\label{GRNT2a7}
\end{eqnarray}
it is said that the covariant derivative and the metric are compatible.  In this regard, there are two main types of  basis \cite{MTW1973}.  Using a coordinate basis, or holonomic basis, we will always have $c_{\mu\nu\alpha} = 0$.  In this case, the connection coefficients are often called Christoffel symbols.  In a noncoordinate basis, or anholonomic basis, we will always have for some of the commutation coefficients $c_{\mu\nu\alpha} \neq 0$. Thus we find that eq. (\ref{GRNT2a7}) will be true for any coordinate basis, but not true for any noncoordinate basis.

%%The coefficients $a_{is}$ and $b^\mu_{is}$ are found \textquotedblleft by the form of the Lie drag of the fields\textquotedblright  \cite{BroB2002}.  

We consider a field free region of space and work with a coordinate basis, then eq. (\ref{GRNT2a2}) reduces to 
\begin{eqnarray}
 2(g^{\mu\beta}\psi_{\mu\alpha} )_{;\beta} = 0,
\label{GRNT2a8a}
\end{eqnarray}
and we find, using the definition eq. (\ref{GRMa2A}) 
\begin{eqnarray}
 G^{\beta}_{\alpha;\beta} = 0.
\label{GRNT2a8}
\end{eqnarray}
This is  the twice-contracted Bianchi Identity, and is always true, being a consequence of the symmetries of the Riemann Tensor.  
%%%%%%%%%%%%%%%%%%%%%%%%%%%%%%%%%%%%%
%%%%%%%%%%%%%%%%%%%%%%%%%%%%%%%%%%%%%%
%
%
%
%
%
\subsection{Noether's Second Theorem and the Response Equation }
%
%
%%%%%%%%%%%%%%%%%%%
%
We start by rearranging terms in eq. (\ref{GRNT2a2})
\begin{eqnarray}
- \psi^{(g)}_{\mu\nu} g^{\mu\nu}_{~;\alpha} + 2(g^{\mu\beta}\psi^{(g)}_{\mu\alpha} )_{;\beta}   = \nonumber \\  \sum_s 
\left[ \psi^s a_{is}   -   \partial_\mu (\psi^s b^\mu_{is}) \right],
\label{GRNT2a9}
\end{eqnarray}
where we have labeled the Euler Expressions for the gravity fields as $(g)$, for future clarity.  We hold the matter fields constant and vary only the metric.  We then find for the RHS
\begin{eqnarray}
  \sum_s \left[ \psi^s a_{is}  -    \partial_\mu (\psi^s b^\mu_{is}) \right] \nonumber \\  \Rightarrow
\psi^{(m)}_{\mu\nu} g^{\mu\nu}_{~;\alpha} - 2(g^{\mu\beta}\psi^{(m)}_{\mu\alpha} )_{;\beta} ,  
\label{GRNT2a10}
\end{eqnarray}
where $\psi^{(m)}_{\mu\nu}$ is the double indexed Euler expression for the matter field when the metric is varied.  
From previous considerations, we have using  eq. (\ref{GRMaxe9})
\begin{eqnarray}
\psi^{(m)}_{\mu\nu} = \frac{\delta \mathscr{L}_{(m)}}{\delta g^{\mu\nu}} = - \frac{\sqrt{-g}}{2} T^{(m)}_{\mu\nu}.
\label{GRNT2a11}
\end{eqnarray}
Subbing eqs. (\ref{GRNT2a10}) and (\ref{GRNT2a11}) into eq. (\ref{GRNT2a9}) and rearranging, we have
\begin{eqnarray}
\psi^{(g)}_{\mu\nu} g^{\mu\nu}_{~;\alpha} + \psi^{(m)}_{\mu\nu} g^{\mu\nu}_{~;\alpha} 
= 2(g^{\mu\beta}\psi^{(g)}_{\mu\alpha} )_{;\beta} \nonumber \\   + 2(g^{\mu\beta}\psi^{(m)}_{\mu\alpha} )_{;\beta}
\label{GRNT2a12}
\end{eqnarray}
In a coordinate basis, the LHS of eq. (\ref{GRNT2a12}) is zero.   By using eqs. (\ref{GRMa2A}) and (\ref{GRNT2a11}) we find
\begin{eqnarray}
    T^{(m)}_{\mu\nu;\beta}  =  G_{\mu\nu;\beta} = 0. 
\label{GRNT2a13}
\end{eqnarray}
This equation, setting the covariant derivative of the stress-energy tensor for the matter fields equal to zero, is called the response equation.  It is considered the response of the matter fields to changes in the gravitational fields.  
%%%%%%%%%%%%%%%%%%%%%%%%%%%%%%%%%%%%%
%%%%%%%%%%%%%%%%%%%%%%%%%%%%%%%%%%%%%%
%
%
%
%
%
\subsection{The Third Theorem }
%
%
%%%%%%%%%%%%%%%%%%%
We will use The Third Theorem to show that the divergence of the Noether current is linked to the Stress-Energy Tensor for the Lagrangian of some fields.  In order for the divergence of one to be zero, it requires the divergence of the other to be zero as well.  This section follows Brown and Brading \cite{BroB2002}.

We assume that our Lagrangian is a function of some fields, $\phi_s$ and the metric.  We start with the second equation of the Third Theorem, eq. (\ref{nNT32})
\begin{eqnarray}
 \psi_s b^\mu_{is} = -  \frac{\partial \mathscr{L}}{\partial(\partial_\mu \phi_s)} a_{is}
  - \partial_\nu \left(\frac{\partial \mathscr{L}}{\partial(\partial_\nu \phi_s)} b^\mu_{is}\right),
\label{GRBTa1}
\end{eqnarray}
where the explicit  sum over $s$ is now implied.  Since we are interested in an infinitesimal coordinate variation, the index $i$ will run over  four, so for consistency we will replace it by $\beta$.   As above, we separate the dependance of the Lagrangian on the fields and the metric on the LHS.  On the RHS we sub in from eq. (\ref{nNT3p5}), but do not expand the second term for brevity, giving, 
\begin{eqnarray}
  2 g^{\mu\gamma} \psi_{\gamma\beta} + \psi_s b^\mu_{\beta s} =  j^\mu_\beta
  - \partial_\nu \left(\frac{\partial \mathscr{L}}{\partial(\partial_\nu \phi_s)} b^\mu_{is}\right),
\label{GRBTa2}
\end{eqnarray}
Now, the first term on the LHS can be written in terms of the stress-energy tensor using eq. (\ref{GRMaxe9}), while, we assume that the Euler expression for the fields will give zero, so we have
\begin{eqnarray}
   \sqrt{-g}  T^\mu _\beta =  j^\mu_\beta
  - \partial_\nu \left(\frac{\partial \mathscr{L}}{\partial(\partial_\nu \phi_s)} b^\mu_{is}\right).
\label{GRBTa3}
\end{eqnarray}
This result agrees with eq. (\ref{GRNT1a5}), and again we see that the stress-energy tensor and the Noether current differs by a divergent term.  We can use the third identity of The Third Theorem to get a relation between the divergence of the current and that of the stress-energy tensor.  The easiest way to see this, is by starting with eq. (\ref{nLgnNTp4}) and equating the $\partial_{\mu\nu}$ terms
\begin{eqnarray}
0 = \frac{\partial \mathscr{L}}{\partial \phi_{ s,\mu}} b^\nu_{is} \partial_{\mu \nu} \Delta p_i .
\label{GRBTa4}
\end{eqnarray}
We integrate this over the volume of application 
\begin{eqnarray}
0 =\int \frac{\partial \mathscr{L}}{\partial \phi_{ s,\mu}} b^\nu_{is} \partial_{\mu \nu} \Delta p_i d^4x.
\label{GRBTa5}
\end{eqnarray}
Next, we integrate by parts twice and use the boundary condition that the functions are zero on the boundary, which gives 
\begin{eqnarray}
0 =\int \partial_{\mu \nu}  \left( \frac{\partial \mathscr{L}}{\partial \phi_{ s,\mu}} b^\nu_{is} \right)  \Delta p_i d^4x,
\label{GRBTa6a}
\end{eqnarray}
or
\begin{eqnarray}
0 = \partial_{\mu \nu}  \left( \frac{\partial \mathscr{L}}{\partial \phi_{ s,\mu}} b^\nu_{is} \right) .
\label{GRBTa6}
\end{eqnarray}
This result is as expected since the term in parenthesis is known to be antisymmetric in $\mu$ and $\nu$. We now take the divergence of eq. (\ref{GRBTa3})
\begin{eqnarray}
   (\sqrt{-g}  T^\mu _\beta)_{,\mu} =  j^\mu_{\beta , \mu}
  - \partial_{\mu\nu} \left(\frac{\partial \mathscr{L}}{\partial(\partial_\nu \phi_s)} b^\mu_{is}\right).
\label{GRBTa7}
\end{eqnarray}
and by eq. (\ref{GRBTa6})
\begin{eqnarray}
   (\sqrt{-g}  T^\mu _\beta)_{,\mu} =  j^\mu_{\beta , \mu}.
\label{GRBTa8}
\end{eqnarray}
We see that the divergence of the Noether current $j^\mu_\beta$ vanishes only when the divergence of $\sqrt{-g}  T^\mu _\beta$ is zero.

%%%%%%%%%%%%%%%%%%%%%%%%%%%%%%%%%%%%%%%%%%%
%%%%%%%%%%%%%%%%%%%%%%%%%%%%%%%%%%%%%%%%%%%
%

 \appendix
 \section{Variable End Points}
\label{appa}
%
%
%%%%%%%%%%%%%%%%%%%%%%%%%%%%%%%%%%%%%%%%%%%
%
In this appendix, we will address the variation of a functional with a variation of the end points of integration.   We will do this for the one dimensional case\cite{GeF1963,MTW1973} and generalise the results for multiple dimensions.\cite{Gol1980}  In the 1-D case, we will use genetic labels in order to focus on the math rather then the implied physics.  Then we generalise the results to apply to Noether's Theorems.  

%%%%%%%%%%%%%%%%%%%%%%%%%%%%%%%%%%%%%%%%%%%%%%
For the general variable end points, we will start with the functional $J$ of some function $f$ of the form 
\begin{eqnarray}
J = 
 {\int_{x_1}^{x_2}
f(y,\dot{y},x)} dx,
\label{VEP1}
\end{eqnarray}
where $y = y(x)$ and $\dot{y} = \frac{\partial y}{\partial x}$.  

\begin{figure*}[h]
\scalebox{0.6}{\includegraphics{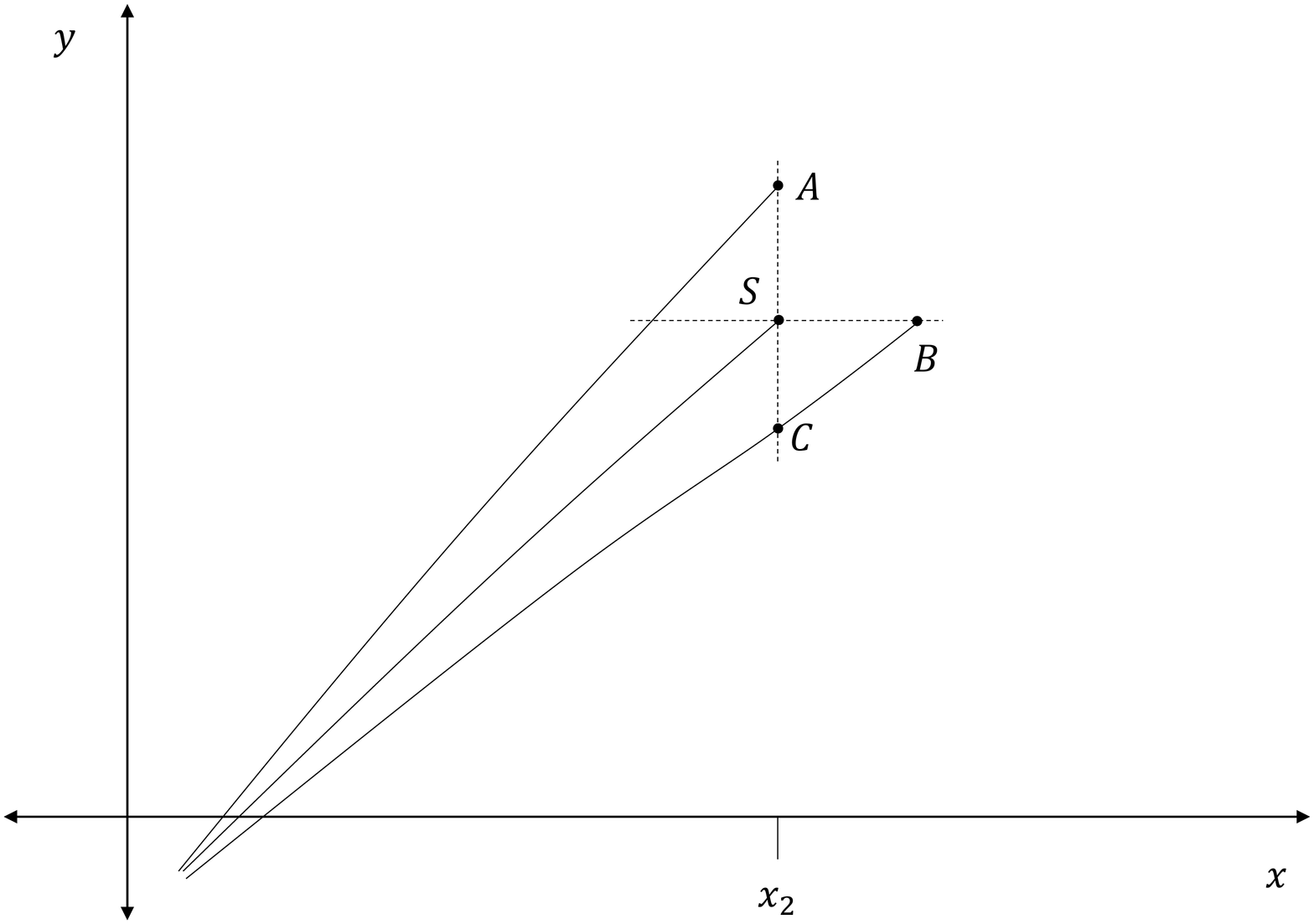}}
\caption{Enlarged area of the variation at $x_2$} 
\label{figdel}
\end{figure*}

We introduce the parameter $\alpha$ in the usual way and thus allowing for the the variation of the end points we have,  $y = y(x, \alpha)$ and $\dot{y} = \dot{y}(x, \alpha)$ and we also have $x = x(\alpha)$.   The only place the variation of x should show up is at the end points of the integration since in between the end points, x is integrated over.   Constructing $\delta J$ in the usual way, we find
\begin{eqnarray}
\delta J = 
 {\int_{x_1}^{x_2}{\left(\frac{\partial f}{\partial y}\delta y 
 +  \frac{\partial f}{\partial \dot{y}}\frac{\partial \dot{y}}{\partial \alpha} d\alpha 
  + \frac{\partial f}{\partial x}\overline{\delta x} \right)
dx}},
\label{Jda5}
\end{eqnarray}
where the $\overline{\delta x}$ is to remind us  that it only has meaning at the end points.
Integrating by parts on the $\delta \dot{y}$ term and recognising that the $\overline{\delta x}$ term can be integrated directly we find
\begin{eqnarray}
\delta J = 
 f \overline{\delta x} \bigg{|}^{x_2 + \Delta}_{x_1 + \Delta}
 + \frac{\partial f}{\partial \dot{y}} \frac{\partial {y}}{\partial \alpha}
 d \alpha \bigg{|}^{x_2 + \Delta}_{x_1 + \Delta} \nonumber \\
 + \int_{x_1}^{x_2}{\left(\frac{\partial f}{\partial y} 
 - \frac{d}{dx} \frac{\partial f}{\partial \dot{y}}\right)\delta y  dx},
\label{JdaG}
\end{eqnarray}
where $\Delta$ is the variation of the end point of integration and we write out $\frac{\partial y}{\partial \alpha} d \alpha$ until we examen the behaviour of the term at the end points.  In order to evaluate the behaviour of the terms at the end points, we refer to Fig.~\ref{figdel}.  Looking at Fig.~\ref{figdel} 
 we see three paths that correspond to three different values of $\alpha$ and terminate at S, A and B.  The path that terminates at S corresponds to $\alpha = 0$.    As $\alpha$ is varied from S along the line $x = x_2$, to point A, we find the usual variation of $y$, thus S-A corresponds to $\delta y$.  As $\alpha$ is varied from S along the line $y = const.$, to point B, we find two contributions to the variation.  The first is the usual variation of $x$, which gives us that S-B corresponds to $\delta x$.  The second is the from the displacement from S to C.  For small values of $\alpha$ we can write this as $- \dot{y} \delta x$.  
So we find, at the end points, that $\overline{\delta x} = \delta x$ and that
 $\partial y / \partial \alpha d \alpha$ leads to $\delta y - \dot{y} \delta x$.  Subbing this into eq.(\ref{JdaG}) gives 
%%%%%%%%%%%%%%%
%
%
%%%%%%%%%%%%%%%%%%
%
%
\begin{eqnarray}
\delta J = 
 \bigg{[} f \delta x 
 + \frac{\partial f}{\partial \dot{y}} (\delta y - \dot{y} \delta x ) \bigg{|}^{x_2 }_{x_1} \nonumber \\
 + \int_{x_1}^{x_2}{\left(\frac{\partial f}{\partial y} 
 - \frac{d}{dx} \frac{\partial f}{\partial \dot{y}}\right)\delta y  dx}.
\label{JdaG2}
\end{eqnarray}
%
%
% %%%%%%%%%%%%%%%%%%%
%
The first term is the contribution from the boundary.  The second term the usual results from a fixed end point variation with the expression in the parenthesis being the Lagrangian Expression, $\psi$.   

For our Lagrangian, we require that the form of the Lagrangian does not change under the transformation, so 
\begin{eqnarray}
\bar{\mathscr{L}}(\bar{\phi}_ s(\bar{x} ^\nu), \bar{\phi} _{s , \mu}(\bar{x}^\nu), \bar{x}^\nu ) 
= \mathscr{L}(\bar{\phi}_ s(\bar{x} ^\nu), \bar{\phi} _{s , \mu}(\bar{x}^\nu), \bar{x}^\nu ) .
\label{Lag12}
\end{eqnarray}
 We can write this variation as
\begin{eqnarray}
\delta I = {\int _{\bar{\Omega}} {\ \mathscr{L}(\bar{\phi}_ s(\bar{x} ^\nu), \bar{\phi} _{s , \mu}(\bar{x}^\nu), \bar{x}^\nu )  d^4\bar{x}_\nu}}  \nonumber \\
-{\int_\Omega {\mathscr{L}(\phi_ s(x ^\nu), \phi _{s , \mu}(x^\nu), x^\nu )  d^4x_\nu}},
\label{Lag2}
\end{eqnarray}
where both the functions $\phi$ and the integration region $\Omega$ can vary.  Eq. (\ref{Lag2}) is not a statement of a extremum but a requirement for a conserved quantity under the variation of some quantity, such as a coordinate or a field function, in some particular way.  

  We start with eq.(\ref{Lag2}) and drop the index $s$ for the moment.  So we have  
\begin{eqnarray}
\delta I = {\int _{\bar{\Omega}} { \mathscr{L}(\bar{\phi}(\bar{x} ^\nu), \bar{\phi} _{, \mu}(\bar{x}^\nu), \bar{x}^\nu )  d^4\bar{x}_\nu}} \nonumber \\
-{\int_\Omega {\mathscr{L}(\phi (x ^\nu), \phi _{, \mu}(x^\nu), x^\nu )  d^4x_\nu}}.
\label{Lag3}
\end{eqnarray}
The bars on  $x^\nu$ can also be dropped since it is the variable of integration.  We can apply eq. (\ref{I3}) to the common area.  Only the difference in the integration region is left.  Thus, we can combine the two integration into one and find,
\begin{eqnarray}
\delta I = {\int_{\Delta \Omega} { \{ \mathscr{L}(\bar{\phi}, \bar{\phi} _{, \mu}, {x}^\nu )  
 - \mathscr{L}(\phi, \phi _{, \mu}, x^\nu )  \} d^4x_\nu}} \nonumber \\
+ \int_\Omega \psi_s \delta \phi_s  
 d^4 x_\nu  ,
\label{Lag31}
\end{eqnarray}
where ${\Delta \Omega}$ is the  boundary contribution, the difference in the integration regions.  To get an expression for this, we will generalise the results above for the 1-D case.    We start with eq.(\ref{JdaG2}) above
\begin{eqnarray}
\delta J =
\bigg{[}\frac{\partial f}{\partial \dot{y}}\delta y
 +(f - \frac{\partial f}{\partial \dot{y}} \dot{y})
 \delta x\bigg{|}_{x1} ^{x_2}
  + \int_{x_1}^{x_2} \psi \delta y  dx.
\label{JG32}
\end{eqnarray}
% %
%
This can be rewritten as 
\begin{eqnarray}
 \delta J = ~~~~~~~~~~~~~~~~~~~~~~~~~~~~~~~~~~~~~~~~~~~~~~~~~~~~~ \nonumber \\ 
 {\int_{x_1} ^{x_2} \Bigg\{ \frac{d}{dx} {\bigg{[}\frac{\partial f}{\partial \dot{y}}\delta y
 +(f - \frac{\partial f}{\partial \dot{y}} \dot{y})
 \delta x\bigg{]} }   + \psi \delta y \Bigg\}  dx},
\label{Lg11}
\end{eqnarray}
 So we have, replacing $f \rightarrow {L}$,  $y \rightarrow \phi$ and $x \rightarrow t$,
\begin{eqnarray}
 \delta J = &&
  \int_{t_1} ^{t_2} \Bigg\{ \frac{d}{dt} {\Bigg[ 
 \frac{\partial {L}}{\partial (\frac{\partial \phi}{\partial t})} \delta \phi 
 + \left(  {L} - \frac{\partial {L}}{\partial (\frac{\partial \phi}{\partial t})} \frac{\partial \phi}{\partial t} \right)
 \delta t \Bigg]} \nonumber \\ &&~~~~~~+ \psi \delta \phi \Bigg\} dt.
\label{Lg12}
\end{eqnarray}
Generalising this for fields we can write for eq.(\ref{Lag31}) as
\begin{eqnarray}
%\begin{split}
\delta I =&& \int_\Omega \Bigg\{    \Bigg[ 
  \left( \mathscr{L} \delta^\mu_\gamma
  -\frac{\partial \mathscr{L}}{\partial  \phi_{,\mu}} { \phi_{, \gamma}} 
 \right)
 \delta x^\gamma 
  +\frac{\partial \mathscr{L}}{\partial \phi_{,\mu}} \delta \phi 
 \Bigg]_{, \mu} \nonumber \\ &&~~~~~~~~+  \psi \delta \phi  \Bigg\}
 d^4 x_\nu.
%\end{split}
\label{Lg13}
\end{eqnarray}
We reintroduce $s$ fields to find
\begin{eqnarray}
\delta I =&& 
\int_\Omega \Bigg\{   \Bigg[ 
  \left(   
   \mathscr{L} \delta^\mu_\gamma
  - \frac{\partial \mathscr{L}}{\partial  \phi_{s,\mu}} { \phi_{s, \gamma}} 
 \right)
 \delta x^\gamma 
  +\frac{\partial \mathscr{L}}{\partial \phi_{s,\mu}} \delta \phi_s 
 \Bigg]_{, \mu} \nonumber \\ &&~~~~~+ \psi_s \delta \phi_s  \Bigg\}
 d^4 x_\nu
= 0,
\label{Lg14}
\end{eqnarray}
which is eq.(\ref{nLg16}).

%%%%%%%%%%%%%%%%%%%%%%%%%%%%%%%%%%%%%%%%%%%%%%%%%%%%%%%%%%%%%%

%%%%%%%%%%%%%%%%%%%%%%%%%%%%%%%%%%

\begin{acknowledgments}
The author would like to thank S. J. A. McLeod for helpful suggestions and for a critical reading of the manuscript.  
\end{acknowledgments}

%%%%%%%%%%%%%%%%%%%%%%%%%%%% 

\bibliography{NTaGR}

\end{document}